%% file: arxiv-unifier.tex
\documentclass[sigconf,nonacm]{acmart}

\usepackage{wrapfig}
\usepackage{subfigure}
\usepackage{makecell}
\usepackage[normalem]{ulem}
\useunder{\uline}{\ul}{}

\input{math_commands}

\input{define-operator}

\AtBeginDocument{%
  \providecommand\BibTeX{{%
    \normalfont B\kern-0.5em{\scshape i\kern-0.25em b}\kern-0.8em\TeX}}}

\begin{document}

\title{Unifie\textsc{r}: A Unified Retriever for Large-Scale Retrieval}


% \affiliation{%

\author{Tao Shen}
\affiliation{%
  \institution{AAII, FEIT, Uni of Technology Sydney}
  % \streetaddress{1 Th{\o}rv{\"a}ld Circle}
  % \city{Hekla}
  \country{Australia}}
\email{tao.shen@uts.edu.au}

\author{Xiubo Geng}
\affiliation{%
  \institution{Microsoft}
  \city{Beijing}
  \country{China}
}
\email{xigeng@microsoft.com}

\author{Chongyang Tao}
\affiliation{%
  \institution{Microsoft}
  \city{Beijing}
  \country{China}
}
\email{chotao@microsoft.com}

\author{Can Xu}
\affiliation{%
  \institution{Microsoft}
  \city{Beijing}
  \country{China}
}
\email{caxu@microsoft.com}

\author{Guodong Long}
\affiliation{%
  \institution{AAII, FEIT, Uni of Technology Sydney}
  % \streetaddress{1 Th{\o}rv{\"a}ld Circle}
  % \city{Hekla}
  \country{Australia}}
\email{guodong.long@uts.edu.au}

\author{Kai Zhang}
\affiliation{%
  \institution{The Ohio State University}
  % \city{Rocquencourt}
  \country{United States}
}
\email{zhang.13253@osu.edu}

\author{Daxin Jiang}
\affiliation{%
  \institution{Microsoft}
  \city{Beijing}
  \country{China}
}
\email{djiang@microsoft.com}

\thanks{Corresponding author: Daxin Jiang (djiang@microsoft.com)}

\renewcommand{\shortauthors}{Shen et al.}

\begin{abstract}
Large-scale retrieval is to recall relevant documents from a huge collection given a query. It relies on representation learning to embed documents and queries into a common semantic encoding space. According to the encoding space, recent retrieval methods based on pre-trained language models (PLM) can be coarsely categorized into either dense-vector or lexicon-based paradigms. These two paradigms unveil the PLMs' representation capability in different granularities, i.e., global sequence-level compression and local word-level contexts, respectively. Inspired by their complementary global-local contextualization and distinct representing views, we propose a new learning framework, Unifie\textsc{r}, which unifies dense-vector and lexicon-based retrieval in one model with a dual-representing capability. Experiments on passage retrieval benchmarks verify its effectiveness in both paradigms. A uni-retrieval scheme is further presented with even better retrieval quality. We lastly evaluate the model on BEIR benchmark to verify its transferability. 
\end{abstract}






\maketitle

\section{Introduction} \label{sec:intro}

Large-scale retrieval aims to efficiently fetch all relevant documents for a given query from a large-scale collection with millions or billions of entries\footnote{An entry can be \textit{passage}, \textit{document}, etc., and we take \textit{document} for demonstrations. }. It plays indispensable roles as a prerequisite for a broad spectrum of downstream tasks, e.g., information retrieval~\cite{Cai2021IRSurvey}, open-domain question answering~\cite{Chen2017DrQA}.
To make online large-scale retrieval possible, the common practice is to represent queries and documents by an encoder in a Siamese manner (i.e., Bi-Encoder, BE) \cite{Reimers2019SentenceBERT}. 
So, its success depends heavily on a powerful encoder by effective representation learning.

Advanced by pre-trained language models (PLM), e.g., BERT~\cite{Devlin2019BERT}, 
recent works propose to learn PLM-based encoders for large-scale retrieval, which are coarsely grouped into two paradigms in light of their encoding spaces with different focuses of representation granularity.
That is, \textit{dense-vector encoding methods} leverage \textit{sequence-level} compressive representations that embedded into dense semantic space \citep{Xiong2021ANCE,Zhan2021STAR-ADORE,Gao2021coCondenser,Khattab2020ColBERT}, whereas \textit{lexicon-based encoding methods} make the best of \textit{word-level} contextual representations by considering either high concurrence~\cite{Nogueira2019DT5Q} or coordinate terms~\cite{Formal2021SPLADE} in PLMs.
To gather the powers of both worlds, some pioneering works propose \textit{hybrid} methods to achieve a sweet point between dense-vector and lexicon-based methods for better retrieval quality. They focus on interactions of predicted scores between the two paradigms.

Nonetheless, such surface interactions -- score aggregations~\cite{Kuzi2020ScoreAgg}, direct co-training~\cite{Gao2021CLEAR}, and logits distillations \cite{Chen2021ImitateSparse} -- cannot fully exploit the benefits of the two paradigms -- regardless of their complementary contextual features and distinct representation views. 
Specifically, as for \textit{contextual features}, the dense-vector models focus more on sequence-level global embeddings against information bottleneck \cite{Lu2021SeedEncoder,Gao2021Condenser,Gao2021coCondenser}, whereas the lexicon-based models focus on word-level local contextual embeddings for precise lexicon-weighting \cite{Formal2021SPLADEv2,Formal2022SPLADE++,Nogueira2019DT5Q}. 
Aligning the two retrieval paradigms more closely is likely to benefit each other since global-local contexts are proven complementary in general representation learning \cite{Shen2019localglobal,Beltagy2020Longformer}.
As for \textit{representing views}, relying on distinct encoding spaces, the two retrieval paradigms are proven to provide different views in terms of query-document relevance \cite{Kuzi2020ScoreAgg,Gao2021CLEAR,Gao2021COIL}. 
Such a sort of `dual views' has been proven pivotal in many previous cooperative learning works \cite{Han2018coteaching,Chen2020HiddenCut,Liang2021rdrop,Gao2021SimCSE}, which provides a great opportunity to bridge the two retrieval paradigms. 
Consequently, without any in-depth interactions, neither the single (dense/lexicon) nor the hybrid retrieval model can be optimal. 

Motivated by the above, we propose a brand-new learning framework, \textbf{Unifie}d \textbf{R}etriever (Unifie\textsc{r}), for in-depth mutual benefits of both dense-vector and lexicon-based retrieval. 
On the one hand, we present a neural encoder with dual representing modules for Unifie\textsc{r}, which is compatible with both retrieval paradigms. 
Built upon an underlying-tied contextualization that empowers consistent semantics sharing, a local-enhanced sequence representation module is presented to learn a dense-vector representation model. Meantime, a global-aware lexicon weighting module considering both the global- and local-context is proposed for a lexicon-based representation. 
On the other hand, we propose a new self-learning strategy, called dual-consistency learning, upon our unified encoder. Besides a basic contrastive learning objective, we first exploit the unified dual representing modules by mining diverse hard negatives for self-adversarial within the Unifie\textsc{r}.  
Furthermore, we present a self-regularization method based on list-wise agreements from the dual views for better consistency and generalization. 

After being trained, Unifie\textsc{r} performs large-scale retrieval via either its lexicon representation by efficient inverted index or dense vectors by parallelizable dot-product. 
Moreover, empowered by our Unifie\textsc{r}, we present a fast yet effective retrieval scheme, \textit{uni-retrieval}, to gather the powers of both worlds, where the lexicon retrieval is followed by a candidate-constrained dense scoring. 
Empirically, we evaluate Unifie\textsc{r} on both passage retrieval benchmarks to check its effectiveness and the BEIR benchmark \cite{Thakur2021BEIR} with twelve datasets (e.g., Natural Questions, HotpotQA) to verify its transferability.

\section{Related Work} \label{sec:rel_work}

\paragraph{PLM-based Retriever.}
Built upon PLMs, recent works propose to learn encoders for large-scale retrieval, which are coarsely grouped into two paradigms in light of their encoding spaces with different focuses of representation granularity:
(i) \textit{Dense-vector encoding methods} directly represent a document/query as a low-dimension sequence-level dense vector $\vu\in\R^e$ ($e$ is embedding size and usually small, e.g., 768). And the relevance score between a document and a query is calculated by dot-product or cosine similarity \cite{Xiong2021ANCE,Zhan2021STAR-ADORE,Gao2021coCondenser,Khattab2020ColBERT}.
(ii) \textit{Lexicon-based encoding methods} make the best of word-level contextualization by considering either high concurrence~\cite{Nogueira2019DT5Q} or coordinate terms~\cite{Formal2021SPLADE} in PLMs. It first weights all vocabulary lexicons for each word of a document/query based on the contexts, leading to a high-dimension sparse vector $\vv\in\R^{|\sV|}$ ($|\sV|$ is the vocabulary size and usually large, e.g., 30k). 
The text is then denoted by aggregating over all the lexicons in a sparse manner. Lastly, the relevance is calculated by lexical-based matching metrics (e.g., BM25~\cite{Robertson2009BM25}).
In contrast, we unify the two paradigms into one sophisticated encoder for better consistency within PLMs, leading to complementary information and superior performance.

\paragraph{Hybrid Retriever.} 
Some works propose to bridge the gap between dense and lexicon for a sweet spot between performance and efficiency. 
A direct method is to aggregate scores of the two paradigms~\cite{Kuzi2020ScoreAgg}, but resulting in standalone learning and sub-optimal quality. 
Similar to our work, CLEAR~\cite{Gao2021CLEAR} uses a dense-vector model to complement the lexicon-based BM25 model, but without feature interactions and sophisticated learning. 
Sharing inspiration with our uni-retrieval scheme, COIL~\cite{Gao2021COIL} equips a simple lexicon-based retrieval with dense operations over word-level contextual embeddings. Unifie\textsc{r} differs in not only our lexicon representations jointly learned for in-depth mutual benefits but also sequence-level dense operations involved for memory-/computation-efficiency. 
Lastly, SPARC~\cite{Lee2020SPARC} distills ranking orders from a lexicon model (BM25) into a dense model as a companion of the original dense vector, which is distinct to our motivation.

\paragraph{Bottleneck-based Learning. } 
In terms of neural designs, our encoder is similar to several recent representation learning works, e.g., SEED-Encoder \cite{Lu2021SeedEncoder}. Condenser \cite{Gao2021Condenser}, coCondenser \cite{Gao2021coCondenser}, and DiffCSE \cite{Chuang2022DiffCSE}, but they focus on the bottleneck of sequence-level dense vectors. 
For example, SEED-Encoder, Condenser, and CoCondenser enhance their dense capabilities by emphasizing the sequence-level bottleneck vector and weakening the word-level language modeling heads, while DiffCSE makes the learned sentence embedding sensitive to the difference between the original sentence and an edited sentence by a word-level discriminator. 
With distinct motivations and targets, we fully exploit both the dense-vector bottleneck and the word-level representation learning in a PLM for their mutual benefits. These are on the basis of not only the shared neural modules but also structure-facilitated self-learning strategies (see the next section). 
Nonetheless, as discussed in our experiments, our model can still benefit from these prior works via parameter initializations.

\paragraph{Instance-dependent Prompt.}
Our model also shares high-level inspiration with recent instance-dependent prompt learning methods \cite{Jin2022InstanceAwarePrompt,Wu2022InstanceDepPrompt}. They introduce a trainable component to generate prompts based on each input example.  Such generated prompts can provide complementary features to the original input for a better prediction quality. 
Analogously, our sequence-level dense vector can be seen as a sort of `soft-prompt' for the sparse lexicon-based representation module, resulting in the superiority of our lexicon-based retrieval, which will be discussed in experiments.
In addition, the `soft-prompt' in our Unifie\textsc{r} also serves as crucial outputs in a unified retrieval system.

\paragraph{Reranker-taught Retriever.}
Distilling the scores from a reranker into a retriever is proven promising~\cite{Hofstatter2020Margin-MSE,Formal2021SPLADEv2,Hofstatter2021TAS-B} . 
In light of this, recent works propose to jointly optimize a retriever and a reranker: RocketQAv2 \cite{Ren2021RocketQAv2} is proposed to achieve their agreements with reranker-filtered hard negatives, while AR2 \cite{Zhang2021AR2} is to learn them in an adversarial fashion where the retriever is regarded as a generator and the reranker as a discriminator. 
In contrast to reranker-retriever co-training, we resort to in-depth sharing from the bottom (i.e., features) to the top (i.e., self-learning) merely within a retriever, with no need for extra overheads of reranker training. 
Meantime, our unified structure also uniquely enables it to learn from more diverse hard negatives mined by its dual representing modules.

\section{Methodology} \label{sec:method}

\paragraph{Task Definition. }
Given a collection with numerous documents (i.e., $\sD = \{d_i\}_{i=1}^{|\sD|}$) and a textual query $q$ from users, a retriever aims to fetch a list of text pieces $\bar{\sD}_q$ to contain all relevant ones. 
Generally, this is based on a relevance score between $q$ and every document $d_i$ in a Siamese manner, i.e., $<\enc(q), \enc(d_i)>$, where $\enc$ is an arbitrary representation model (e.g., Bag-of-Words and neural encoders) and $<\cdot,\cdot>$ denotes a lightweight relevance metric (e.g., BM25 and dot-product).

\subsection{General Retriever Learning Framework}

\begin{wrapfigure}{R}{0.22\textwidth}
% \begin{figure}[t]
% \vspace{-10pt}
    \centering
    \includegraphics[width=0.22\textwidth]{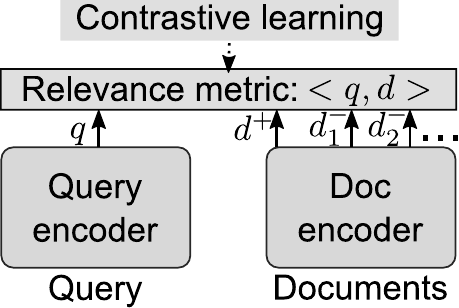}
    % \vspace{-5pt}
    \caption{\small Bi-encoder learning. }
    \label{fig:model_contrast}
    \vspace{-10pt}
% \end{figure}
% \vspace{-6mm}
\end{wrapfigure}

To ground a method, we first introduce a contrastive learning framework to train a retrieval model (Figure~\ref{fig:model_contrast}). 
For supervision data in retriever training, differing from traditional categorical tasks, only query-document tuples (i.e., $(q, d^+_q)$) are given as positive pairs. 
Hence, given a $q$, a method needs to sample a set of negatives $\sN_q=\{d^-_q\}_{1}^M$ from $\sD$, and trains the retriever on tuples of $(q, d^+_q, \sN_q)$. $M$ is the number of negatives. 
If no confusion is caused, we omit the subscript `$q$' for a specific query in the remaining. 

Formally, given $q$ and $\forall d \in\{d^+\} \cup \sN$, an encoder, $\enc(\cdot; \theta)$, is applied to them individually to produce their embeddings, i.e., $\enc(q; \theta)$ and $\enc(d; \theta)$,
where the encoder is parameterized by $\theta$ if applicable.
It is noteworthy we tie the query encoder with the document encoder in our work for simplicity. Then, a relevance metric is applied to each pair of the embeddings of the query and each document. Thus, a probability distribution over the documents $\{d^+\}\!\cup\!\sN$ can be defined as
\begin{align}
    &\vp \coloneqq P(\rd~|~q, \{d^+\}\!\cup\!\sN; \theta) =  \label{equ:gene_prob_dist} \\
    \notag &~~\dfrac{\exp(<\enc(q; \theta), \enc(d; \theta)>)}{\sum_{d'\in\{d^+\} \cup \sN} \exp(<\enc(q; \theta), \enc(d'; \theta)>)},
\end{align}
where $\forall d\in\{d^+\} \cup \sN$. Lastly, a contrastive learning loss to optimize the encoder $\theta$ is
\begin{align}
    L_{\theta} = - \log P(\rd=d^+~|~q, \{d^+\}\!\cup\!\sN; \theta). \label{equ:gene_contrast}
\end{align}

\subsection{Neural Encoder in Unifie\textsc{r}} \label{sec:method_encoder}

\begin{figure}[t]
\centering
\includegraphics[width=0.75\linewidth]{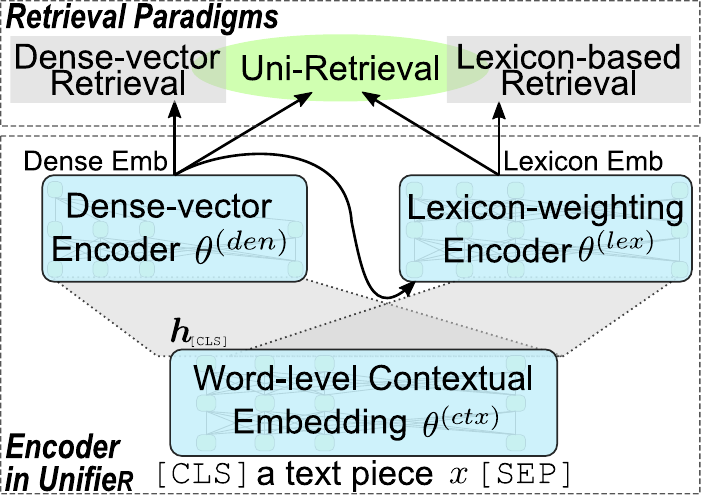}
\caption{\small The encoder in Unifie\textsc{r}.
}\label{fig:model_encoder}
\end{figure}

We present an encoder (see Figure~\ref{fig:model_encoder}) for Unifie\textsc{r} for dense-vector and lexicon-based retrieval. 

\paragraph{Underlying-tied Contextualization. } 
We first propose to share the low-level textual feature extractor between both representing paradigms. 
Although the two paradigms are focused on different representation granularities, sharing their underlying contextualization module can still facilitate semantic knowledge transfer between the two paradigms. 
As such, they can learn consistent semantic and syntactic knowledge towards the same retrieval targets, especially the salient lexicon-based features transferred to dense vectors. 
Formally, we leverage a multi-layer Transformer \cite{Vaswani2017Transformers} encoder to produce word-level (token-level) contextualized embeddings, i.e., 
\begin{align}
    \mH^{(x)} = \transformerenc(\texttt{[CLS]} x \texttt{[SEP]};\theta^{(ctx)})
    \label{equ:encoder_ctx}
\end{align}
where $\forall x \in \{q\} \cup \{d^+\} \cup \sN$, and \texttt{[CLS]} \& \texttt{[SEP]} are special tokens by following PLMs \cite{Devlin2019BERT,Liu2019RoBERTa}, $\mH^{(x)}=[\vh_\texttt{[CLS]}^{(x)}, \vh_1^{(x)}, \dots, \vh_n^{(x)}, \vh_\texttt{[SEP]}^{(x)}]$ are resulting embeddings, and $n$ is the number of words in $x$.

\paragraph{Local-enhanced Sequence Representation.}
On top of the embeddings with enhanced local contexts, we then present a representing module to produce sequence-level dense vectors. 
For this purpose, we apply another multi-layer Transformer encoder to $\mH^{(x)}$, followed by a pooler to derive a sequence-level vector. This can be written as
\begin{align}
    \vu^{(x)}  = \pool(\transformerenc(\mH^{(x)}; \theta^{(den)})), \label{equ:encoder_dense}
\end{align}
where this module is parameterized by $\theta^{(den)}$ untied with $\theta^{(ctx)}$, $\pool(\cdot)$ gets a sequence-level dense vector by taking the embedding of special token \texttt{[CLS]}, and the resulting $\vu^{(x)}\in\R^e$ denotes a global dense representation of the input text $x$, which is used for dense-vector retrieval. 

\paragraph{Global-aware Lexicon Weighting.}
Lastly, to achieve lexicon-based retrieval, we adapt a recent SParse Lexical AnD Expansion Model (SPLADE) \cite{Formal2021SPLADEv2} into our neural encoder. 
SPLADE is a lexicon-weighting retrieval model which learns sparse expansion for each word in query/document $x$ via the MLM head of PLMs and sparse regularization.
Differing from the original SPLADE, our lexicon-based representing module not only shares its underlying feature extractor with a dense model but strengthens its hidden states by the global vector $\vu^{(x)}$ above. 
The intuition is that, similar to text decoding with a bottleneck hidden state, the global context serves as high-level constraints (e.g., concepts/topics) to guide word-level operations \cite{Sutskever2014seq2seq,Lu2021SeedEncoder,Gao2021Condenser}.
In particular, the word-level contextualization embeddings passed into this module are manipulated as $\hat\mH^{(x)}=[\vu^{(x)}, \vh_1^{(x)}, \dots, \vh_\texttt{[SEP]}^{(x)}]$. 
Then, a lexicon-weighting representation for $x$ can be derived by 
\begin{align}
    \vv^{(x)} \!\!=\! \log(& 1 \!+\!  \maxpool(\relu( \label{equ:encoder_lexicon}\\
    \notag &\mW^{(e)}\transformerenc(\hat\mH^{(x)}); \theta^{(mlm)}))), 
\end{align}
where, $\theta^{(mlm)}$ parameterizes a multi-layer Transformer encoder, $\mW^{(e)}\in\R^{|\sV|\times e}$ denotes the transpose of word embedding matrix as the MLM head, $|\sV|$ denotes the vocabulary size, $\theta^{(lex)} = \{\mW^{(e)}, \theta^{(mlm)}\}$ parameterizes this module, and $\vv^{(x)}\in\R^{|\sV|}$ is a sparse lexicon-based representation of $x$. And its sparsity is regularized by FLOPS \cite{Paria2020FLOPS} as in \cite{Formal2021SPLADEv2}. Here, the saturation function $\log(1+\maxpool(\cdot))$ prevents some terms from dominating. 

In summary, given a text $x$, Unifie\textsc{r} produces two embeddings via its dual representing modules:
\begin{align}
    \notag \vu^{(x)} \coloneqq \denseenc(x; \Theta^{(den)}), \\
    \vv^{(x)} \coloneqq \lexiconenc(x; \Theta^{(lex)}),
    \label{equ:define_params}
\end{align}
where $\Theta^{(den)}\!=\! \{\theta^{(ctx)},\! \theta^{(den)}\}~\text{and}~\Theta^{(lex)}\!=\!\{\theta^{(ctx)},\! \theta^{(den)},\! \theta^{(lex)}\}.$
Hence, $\vu^{(x)}\in\R^e$ denotes a dense vector and $\vv^{(x)}\in\R^{|\sV|}$ denotes a sparse lexicon-based embedding.

\begin{figure}[t]
\centering
\includegraphics[width=\linewidth]{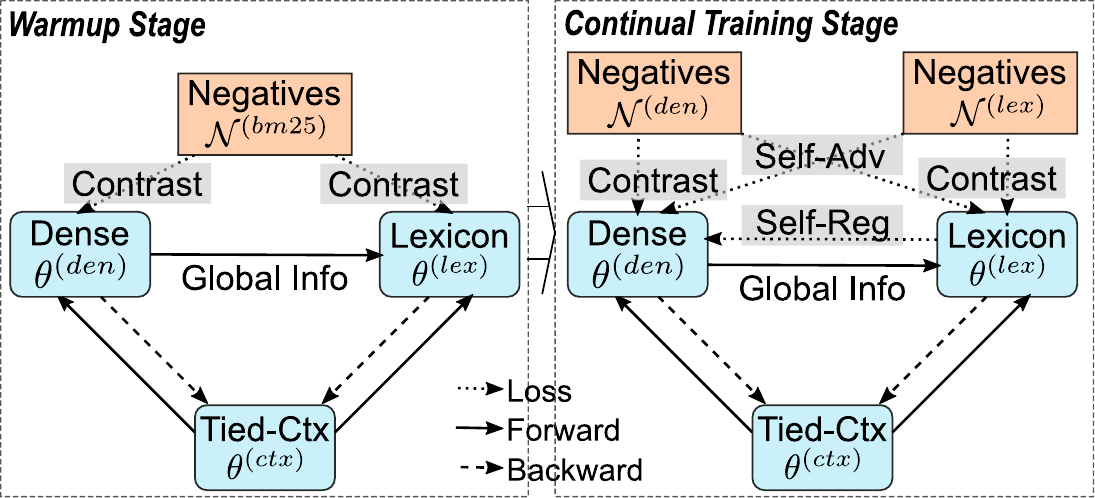}
\caption{\small The two-stage self-learning strategy for Unifie\textsc{r}.}
\label{fig:model_learning}
\end{figure}

\subsection{Dual-Consistency Learning for Unifie\textsc{r}} \label{sec:method_learning}
To maximize our encoder's representing capacity, we propose a self-learning strategy, called dual-consistency learning (Figure~\ref{fig:model_learning}). The `dual-consistency' denotes learning the dual representing modules to achieve consistency in a unified model via \textit{negative samples} and \textit{module predictions}.

\paragraph{Basic Training Objective. }
To learn the encoder, a straightforward way is applying the contrastive learning loss defined in Eq.(\ref{equ:gene_prob_dist}-\ref{equ:gene_contrast}) to our dual representing modules. 
That is,
\begin{align}
    \notag L^{\text{(con)}} &= - \log P(\rd = d^+|q, \{d^+\}\!\cup\!\sN; \Theta^{(den)}) \\
    &- \log P(\rd\!=\!d^+|q, \{d^+\}\!\cup\!\sN; \Theta^{(lex)}), \label{equ:foundation_loss}
\end{align}
where the former is for dense-vector retrieval while the latter is for lexicon-based retrieval. 
Towards the same retrieval target, the model is prone to learn consistent semantic and syntactic features via complementing the global-local granularity of the two retrieval paradigms. 
Due to the non-differentiability of lexicon-based metrics, we follow \citep{Formal2021SPLADE} to use dot-product of lexicon-weighting representation during training but resort to a lexicon matching system \cite{Yang2017Anserini} with quantization during indexing\&retrieval. (see Appx.~\ref{sec:retrieval_scheme} for details)
Note that 
$\theta^{(den)}$ would not be optimized w.r.t. the losses on top of the lexicon-based module. 
As for the query's negatives $\sN$ of in Eq.(\ref{equ:foundation_loss}), they are initially sampled by a BM25 retrieval system at \textit{the warmup stage} \cite{Zhan2021STAR-ADORE,Gao2021coCondenser}, denoted as
% \begin{align}
    $\sN^{(bm25)} = \left\{d|d \sim P(\rd~|~q, \sD_{\backslash\{d^+\}}; \text{BM25})\right\}$,
% \end{align}
where $\sD_{\backslash\{d^+\}}$ denotes all documents in the collection $\sD$ except the positive $d^+$ for query $q$.

\input{tables/main_dev_update1}

\paragraph{Negative-bridged Self-Adversarial. } However, it is verified that learning a retriever based solely on BM25 negatives cannot perform competitively \cite{Xiong2021ANCE,Zhan2021STAR-ADORE}. 
Thereby, previous works propose to sample hard negatives by the best-so-far retriever for continual training \cite{Zhan2021STAR-ADORE,Gao2021coCondenser}, a.k.a. self-adversarial learning \cite{Sun2019RotatE}.
In our pilot experiments, we found the two retrieval paradigms can provide distinct hard negatives ($>40\%$ top-retrieved candidates are different) to ensure diversity after a combination. 
This motivates us to make the best of the hard negatives sampled by our dual representing modules: hard negatives sampled from one module can be applied to both itself and its counterpart in one unified framework. 
This can be regarded as a sort of self-distillation as both distilling samples (i.e., document mined from the collection) and distilling labels (i.e., negative label only) are sourced from one unified model. 
So, we first sample two sets of negatives from the dual-representing modules:
\begin{align}
    \notag \sN^{(den)} \!=\! \left\{d|d \sim P(\rd~|~q, \sD_{\backslash\{d^+\}}; \Theta^{(den)} )\right\}, \\
    \sN^{(lex)} \!=\! \left\{d|d \sim P(\rd~|~q, \sD_{\backslash\{d^+\}}; \Theta^{(lex)} )\right\}, \label{equ:hard_neg_sample}
\end{align}
where our Unifie\textsc{r} was trained with $\sN^{(bm25)}$ at \textit{warmup stage}. 
Next, we upgrade $\sN$ in Eq.(\ref{equ:foundation_loss}) from $\sN^{(bm25)}$ at \textit{warmup stage} to $\sN^{(den)} \cup \sN^{(lex)}$, and then perform a \textit{continual learning stage}.

\paragraph{Agreement-based Self-Regularization. }
We lastly present a self-regularization method for Unifie\textsc{r}. 
Its goal is to achieve an agreement from different views through our dual representing modules. 
Such an agreement-based self-regularization has been proven effective in both retrieval model training (via retriever-reranker agreements for consistent results \cite{Ren2021RocketQAv2,Zhang2021AR2}) and general representation learning (via agreements from various perturbation-based views for better generalization \cite{Chen2020HiddenCut,Liang2021rdrop,Gao2021SimCSE}). 
It is stronger than the contrastive learning in Eq.(\ref{equ:foundation_loss}) as the agreement is learned by a KL divergence, i.e., \begin{align}
    L^{\text{(reg)}} \!=\! \KL (P(\rd|q, &\{d^+\}\!\cup\!\sN; \Theta^{(den)})  \label{equ:kl_loss}\\ 
    \notag &\Vert P(\rd|q, \{d^+\}\!\cup\!\sN; \Theta^{(lex)})).%\label{equ:kl_loss}
\end{align}

\paragraph{Overall Training Pipeline.}
In line with \cite{Gao2021coCondenser}, we lastly follow a simple three-step pipeline to learn our retriever on the basis of the proposed training objectives and hard negatives:
    (i) \textbf{\textit{Warmup Stage}}: Initialized by a pre-trained model, Unifie\textsc{r} is updated w.r.t. Eq.(\ref{equ:foundation_loss}) + $\lambda~\text{FLOPS}$ (by following \cite{Formal2021SPLADEv2} for sparsity), with BM25 negatives $\sN^{(bm25)}$. % by Eq.(\ref{equ:hn_bm25}).
% \vspace{-4pt}
    (ii) \textbf{\textit{Hard Negative Mining}}: According to the warmup-ed Unifie\textsc{r}, static hard negatives, $\sN^{(den)}$ and $\sN^{(lex)}$, are sampled by Eq.(\ref{equ:hard_neg_sample}).
% \vspace{-4pt}
    (iii) \textbf{\textit{Continual Learning Stage}}: Continual with the warmup-ed  Unifie\textsc{r}, the model is finally optimized on $\sN^{(den)} \cup \sN^{(lex)}$ w.r.t. a direct addition of Eq.(\ref{equ:foundation_loss}\&\ref{equ:kl_loss})+$\lambda~\text{FLOPS}$.
% \end{enumerate}

\subsection{Retrieval Schemes} \label{sec:retrieval_scheme}

\paragraph{Inference of Lexicon-based Retrieval.} During the inference of large-scale retrieval, there are some differences between dense-vector and lexicon-based retrieval methods. 
As in Eq.(\ref{equ:gene_prob_dist}), we use the dot-product between the real-valved sparse lexicon-based representations as a relevance metric, where `real-valved' is a prerequisite of gradient back-propagation and end-to-end learning.  
However, it is inefficient and infeasible to leverage the real-valved sparse representations, especially for the open-source term-based retrieval systems, e.g., LUCENE and Anserini \citep{Yang2017Anserini}. Following \citet{Formal2021SPLADEv2}, we adopt `quantization' and `term-based system' to complete our retrieval procedure. That is, to transfer the high-dimensional sparse vectors back to the corresponding lexicons and their virtual frequencies, the lexicons are first obtained by keeping the non-zero elements in a high-dim sparse vector, and each virtual frequency then is derived from a straightforward quantization (i.e., $\floor{100\times\vv}$). In summary, the overall procedure of our large-scale retrieval based on a fine-tuned Unifie\textsc{r}-lex is i) generating the high-dim sparse vector for each document and transferring it to lexicons and frequencies, ii) building a term-based inverted index via Anserini \citep{Yang2017Anserini} for all documents in a collection, iii) given a test query, generating the lexicons and frequencies, in the same way, and iv) querying the built index to get top document candidates.

\paragraph{Uni-retrieval Scheme.} As in Figure~\ref{fig:model_encoder}, our model is fully compatible with the previous two retrieval paradigms. 
In addition, we present a \textit{uni-retrieval} scheme for fast yet effective large-scale retrieval. 
Instead of adding their scores \cite{Kuzi2020ScoreAgg,Formal2022SPLADE++} from twice-retrieval with heavy overheads, we pipelinelize the retrieval procedure: given $q$, our lexicon-based retrieval under an inverted file system is to retrieve top-K documents from $\sD$. Then, our dense-vector retrieval is then applied to the constrained candidates for dense scores. The final retrieval results are according to a simple addition of the two scores. 
We use `addition' as our combination baseline for its generality and explore more advanced methods in \S\ref{sec:advanced_arc}. 
And, due to fast dense-vector dot-product calculations on top-K documents, uni-retrieval's latency is almost equal to single lexicon-based retrieval.

% \begin{wraptable}{r}{0.38\textwidth} 
\begin{table}[t]
\centering
% \vspace{-35pt}
% \resizebox{0.38\textwidth}{!}{
    % \setlength{\tabcolsep}{1.5pt}
    \small
    \caption{\small MS-Marco retrieval on \textit{one-positive-enough} recall.} \label{tab:exp_main_otherrecall}
    \begin{tabular}{lccc}
    \toprule
    \multicolumn{1}{c}{\textbf{Method}} & {M@10}        & \underline{\textit{R@50}}         & \underline{\textit{R@1K}}          \\ \hline
    RocketQA \cite{Qu2021RocketQA}        & 37.0  & 85.5    & 97.9  \\
    PAIR \cite{Ren2021PAIR}    & 37.9& 86.4     & 98.2   \\
    RocketQAv2   & 38.8    & 86.2    & 98.1    \\
    AR2 & 39.5    & 87.8 & \textbf{98.6} \\ \hline
    Unifie\textsc{r}$_{\text{lexicon}}$  & 39.7   & 87.6 & 98.2 \\
    Unifie\textsc{r}$_{\text{dense}}$   & 38.8    & 86.3  & 97.8 \\ 
    Unifie\textsc{r}$_{\text{uni-retrieval}}$   & \textbf{40.7} & \textbf{88.2}   & 98.5    \\
    \bottomrule
\end{tabular}
% \vspace{-5pt}
% \vspace{-8pt}
% }
% \vspace{-15pt}
% \end{wraptable}
\end{table}

\section{Experiment} \label{sec:exp}
\paragraph{Datasets \& Metrics.} 
In line with \citep{Formal2021SPLADEv2}, we use popular passage retrieval datasets, MS-Marco~\cite{Nguyen2016MSMARCO}, with official queries (no augmentations \cite{Ren2021RocketQAv2}), and report for MS-Marco Dev set and TREC Deep Learning 2019 set~\cite{Craswell2020TREC19}. 
Following previous works, we report MRR@10 (M@10) and Recall@1/50/100/1K\footnote{
We follow official evaluation metrics at
% \url{https://github.com/usnistgov/trec_eval} and 
\url{https://github.com/castorini/anserini}. 
But, we found 2 kinds of Recall@N on MS-Marco in recent papers, i.e., official \textit{all-positive-macro recall} and \textit{one-positive-enough recall} (see \S\ref{sec:recall_metric} for details). Thereby, we report the former by default but list the latter separately for fair comparisons.} for MS-Marco Dev, and report nDCG@10 and R@100 for TREC Deep Learning 2019. 
Besides, we also transfer our model trained on MS-Marco to the BEIR benchmark~\cite{Thakur2021BEIR} to evaluate its generalizability, where nDCG@10 is reported. 
We take 12 datasets (i.e., TREC-COVID, NFCorpus, NQ, HotpotQA, FiQA, ArguAna, Tóuche-2020, DBPedia, Scidocs, Fever, Climate-FEVER, and SciFact) in the BEIR benchmark as they are widely-used across most previous papers.

\paragraph{Experimental Setups.}
% \label{sec:exp_setups}
As stated in \S\ref{sec:method_learning}, we take a 2-stage learning scheme \cite{Gao2021coCondenser}.
We use coCondenser-marco \cite{Gao2021coCondenser} (unsupervised continual pre-training from BERT-base \cite{Devlin2019BERT}) as our initialization as it shares a similar neural structure and has potential for promising performance \cite{Gao2021coCondenser,Formal2022SPLADE++,Zhang2021AR2}. 
$\theta^{(ctx)}$, $\theta^{(den)}$, and $\theta^{(lex)}$ correspond to Transformer layers of 6, 6, and 2, respectively, where max length is 128 and warmup ratio is 5\%. 
At warmup stage, batch size of queries is 16, each with 1 positive document and 15 negatives, learning rate is $2\!\times\!10^{-5}$, the random seed is fixed to 42. And loss weight of FLOPS \cite{Paria2020FLOPS} is set to 0.0016 since we want make the model sparser than SPLADE \cite{Formal2021SPLADEv2} (0.0008). 
At continual learning stage, batch size is 12 to enable each module with 15 negatives. And learning rate is reduced to 1/3 of the original, and the random seed is changed to 22 for a new data feeding order. And the loss weight of FLOPS is lifted to 0.0024. We did not tune the hyperparameters. 
In retrieval phase, we set K=2048 in our uni-retrieval, and also compare other choices in our analysis. 
All experiments are run on a single A100 GPU. Our codes are released at \url{https://github.com/taoshen58/UnifieR}.
% Please refer to \S\ref{sec:exp_setups} for our pre-training and fine-tuning setups. 

\subsection{Main Evaluation}

\paragraph{MS-Marco Dev.} As in Table~\ref{tab:exp_main_dev_dl19}\&\ref{tab:exp_main_otherrecall}, our framework achieves new state-of-the-art metrics on most metrics. 
Our dense-vector retrieval surpasses previous methods without distillations from rerankers, while our lexicon-based retrieval pushes the best sparse method to a new level, especially in MRR@10 (+1.4\%). Empowered by our unified structure, the uni-retrieval scheme can achieve 40.7\% MRR@10.
Although R@1K is approaching its ceiling across recent works, we notice Unifie\textsc{r} is less competitive than AR2 (-0.2\%) in Table~\ref{tab:exp_main_otherrecall}, as the latter involves a costly reranker in training for better generalization. And please see \S\ref{sec:advanced_arc} for our rerank-taught results. 

\paragraph{TREC Deep Learning 2019.}
% Table 3
As listed in Table~\ref{tab:exp_main_dev_dl19}, our retrieval method, with either single (dese/lexicon) or unified representation, achieves a state-the-of-art or competitive retrieval quality. 
Specifically, compared to the previous best method, called TAS-B, our model lifts MRR@10 and nDCG@10 by 6.9\% and 2.6\%, respectively.

\begin{table}
% \begin{wraptable}{R}{0.52\textwidth}  
\small
\centering
% \vspace{-22pt}
% \setlength{\tabcolsep}{0.5pt}
\caption{\small Retrieval nDCG@10 results on BEIR with 12 out-of-domain datasets, and 1 in-domain dataset. \textbf{Avg} is mean nDCG over 12 datasets and \textbf{Best} is how many datasets a method achieves best. DocT5Query \cite{Nogueira2019DT5Q}, ColBERT \cite{Khattab2020ColBERT}, ColBERT-v2 \cite{Khattab2021ColBERTv2}, DistilSPLADE \cite{Formal2021SPLADEv2}.}
\label{tab:exp_main_beir}
\begin{tabular}{clcc|c}
\toprule
\multicolumn{2}{c}{\textbf{Method}} & \textbf{Avg} & \textbf{Best} & \textbf{In-Dm}  \\
\hline
\multirow{3}{*}{\makecell{Lexicon \\-based}} & BM25 \cite{Thakur2021BEIR} & 41.1 & 1 & 22.8 \\
 & DocT5Query & 42.4& 0 & 33.8 \\
  & UniCOIL \citep{Lin2021UniCOIL} & 40.0& 0 & - \\
 \hline
\multirow{5}{*}{\makecell{Dense \\ -vector}} & ColBERT & 41.8& 2 & 40.8 \\ 
& ANCE \cite{Xiong2021ANCE} & 37.7 & 0 & 38.8 \\
 & GenQ \cite{Thakur2021BEIR} & 39.8 & 1 & 40.8 \\
  & TAS-B \cite{Hofstatter2021TAS-B} & 40.4 & 0 & 40.8 \\
   & Contriever \cite{Izacard2021Contriever} & 44.3 & \textbf{4} & - \\ 
   \hline
\multicolumn{2}{l}{~~~~~~~~~~~~~~~~Unifie\textsc{r}$_{\text{uni-retrieval}}$} & \textbf{44.5} & \textbf{4}  & \textbf{47.1} \\
\hline\hline
\multirow{2}{*}{\makecell{Reranker \\ taught}} & ColBERT-v2  & 47.0 & N/A & 42.5 \\
 & DistilSPLADE & 47.0 & N/A & 43.3 \\
   \hline
\multirow{2}{*}{\makecell{Huge \\ models}} & GTR-XXL \cite{Ni2021GTR} & 45.9 & N/A  & 44.2 \\
 & SGPT-5.8B \cite{Muennighoff2022SGPT} & 49.4 & N/A  & 39.9 \\
\bottomrule
\end{tabular}
% \vspace{-5pt}
% \vspace{-12pt}
\end{table}
% \end{wraptable}
\paragraph{BEIR Benchmark.}
Table~\ref{tab:exp_main_beir} shows in-domain evaluation and zero-shot transfer on BEIR  (see \S\ref{sec:beir_full}). 
It is observed that, with outstanding in-domain inference ability, our model also delivers comparable transferability among the retrievers with similar training settings (i.e., comparable models o/w reranker distillations). 
But, we found our model suffers from inferior generalization ability compared to the models with MSE-based reranker distillation \cite{Khattab2021ColBERTv2,Formal2021SPLADEv2}. 
And a small model with distillation (e.g., DistilSPLADE) even beats the models with billions of parameters (e.g., GTR-XXL). 
The potential reasons are two-fold: i) distilling a reranker to the retriever has been proven to produce more generalizable scores than a bi-encoder \cite{menon2021indefense} and ii) the initialization of Unifie\textsc{r}, coCondenser, has been pre-trained on Marco collection, reducing its generalization.

\subsection{Further Analysis}

% \begin{wraptable}{R}{0.36\textwidth} \small
\begin{table}[t] \small \centering
% \vspace{-18pt}
% \vspace{3pt}
% \setlength{\tabcolsep}{2pt}
\caption{\small Comparison with ensemble and hybrid retrievers. $^1$We operate on the best SPLADE model (MRR@10=38.5) with the best coCondenser (MRR@10=38.2). $^2$An ensemble of four SPLADE models. }\label{tab:exp_quan_ensemble}
\begin{tabular}{lcc}
\toprule
\multicolumn{1}{l}{\textbf{Method}} & {M@10} & {R@1} \\
\hline
Unifie\textsc{r}$_{\text{uni-retrieval}}$ & 40.7 & 26.9 \\
\hline
Uni-scheme of Best$^1$ & 40.3 & 26.1 \\
Ensemble of Best$^1$ & 40.4 & 26.5 \\
Ensemble of SPLADE$^2$ & 40.0 & - \\
COIL-full (hybrid) & 35.5 & - \\
\bottomrule
\end{tabular}
% \vspace{-5pt}
% \end{table}
% \vspace{-18pt}
% \end{wraptable}
\end{table}

\paragraph{Comparison to Ensemble Models.}
As in Table~\ref{tab:exp_quan_ensemble}, we report the numbers to compare our uni-retrieval scheme with ensemble models. 
Even if we only need once large-scale retrieval followed by a small amount of dot-product calculation, the model still surpasses its competitors. 
Meantime, we found both uni-retrieval and ensemble are bounded by the worse participant. 
For example, even if we use a SPLADE with MRR@10 of 39.3 for `Ensemble/Uni-scheme of Best', the performance did not show a remarkable gain. This suggests us to look for a better aggregation method in the future.

\input{tables/ablation}
\paragraph{Ablation of Neural Structure.}
To verify the correctness of each module design, we conduct an ablation study on the neural structure of the encoder (\S\ref{sec:method_encoder}) in Table~\ref{tab:exp_abla_encoder}. 
This must be performed at the warmup step as the second stage is continual from the warmup. 
It is observed that, either removing the global information from the lexicon-based module or discarding in-depth inter-paradigm interactions (i.e., learning independently) degrades the model dramatically. 
Surprisingly, removing the global also diminishes dense performance. A potential reason is that, such a change makes the fine-tuning inconsistent with its initializing pre-trained model, coCondenser, leading to corrupted representing capability. 
% Please refer to \S\ref{sec:alba_learn} for additional ablations about learning and data. 

% \subsection{Ablation of Learning Objectives} \label{sec:alba_learn}

\begin{table}[t]\small
\centering
\caption{\small Learning strategy for \textit{continual training} on MS-Marco Dev.}
\label{tab:exp_abla_learning}
\begin{tabular}{lcccc}
\toprule
\multirow{2}{*}{{\textbf{Methods}}} &  \multicolumn{2}{c}{{Lexicon-based}} & \multicolumn{2}{c}{{Dense-vector}} \\
% \cline{2-3} \cline{4-5}
\cmidrule(lr){2-3} \cmidrule(l){4-5}
& {M@10} & {R@100}  & {M@10} & {R@100} \\
\hline
Unifie\textsc{r} & 39.7 & 91.2 & 38.8 & 90.3 \\
~$\Diamond$ w/o Self-adv & 39.6 & 91.5 & 38.2 & 90.3 \\
~$\Diamond$ w/o Self-adv\&-reg & 39.5 & 91.3 & 37.9 & 90.1 \\
\bottomrule
\end{tabular}
\end{table}

\paragraph{Study on Learning Strategy.}
Furthermore, we conduct another study on the learning strategies (\S\ref{sec:method_learning}) in Table~\ref{tab:exp_abla_learning}. This is performed at the continual training stage. 
The table shows that, ablating the negative-bridged self-adversarial (self-adv) and the agreement-based self-regularization (self-reg) has a minor effect on lexicon-based retrieval but is remarkable on dense-vector one. This is because the former is already far stronger than the latter. Thereby, both self-adv and self-reg can be regarded as a sort of (self-)distillation from lexicon knowledge from a well-trained language model to dense semantic representation. We will dive into the self-reg in the following to seek for a better learning strategy, especially for the lexicon-based retrieval. 
In addition, we also observed that the proposed self-learning strategies (i.e., self-adversarial and self-regularization) mainly contribute to dense-vector retrieval (+0.6\% and 0.3\% MRR@10, respectively) but only bring limited performance improvement for lexicon-based method (+0.1\% and 0.1\% MRR@10, respectively). The main reasons are two-fold: 
i) Verified in \citep{Formal2021SPLADEv2,Hofstatter2021TAS-B}, lexicon-based methods consistently outperform dense-vector methods in ad-hoc retrieval as lexicon-overlap serves as an important feature in relevance calculations. Therefore, the improvement mainly falls into the dense-vector part via knowledge distillation from the lexicon-based part.
ii) Meantime, the common knowledge distillation schema is from a strong teacher to a weak student, e.g., cross-encoder reranker v.s. bi-encoder retriever with a 5$\sim$10\% performance gap in ad-hoc retrieval scenarios \citep{Zhang2021AR2,Ren2021RocketQAv2}. In contrast, the participants (Unifie\textsc{r}-dense \& -lexicon) of our self-learning have similar performance (gap <1\%), making the improvement limited.

\begin{table}[t]
\small
\centering
\caption{\small Effect of our self-regularization's targets on MS-Marco.}
\label{tab:exp_dive_self_reg}
\begin{tabular}{lcccc}
\toprule
\multirow{2}{*}{{\textbf{Methods}}} &  \multicolumn{2}{c}{{Lexicon-based}} & \multicolumn{2}{c}{{Dense-vector}} \\
% \cline{2-3} \cline{4-5}
\cmidrule(lr){2-3} \cmidrule(l){4-5}
& {M@10} & {R@100}  & {M@10} & {R@100} \\
\hline
Unifie\textsc{r} & 39.7 & 91.2 & 38.8 & 90.3 \\
~$\Diamond$ Self-reg on $\sN^{(den)}$ only  & 39.5 & 91.0 & 38.3 & 90.0 \\
~$\Diamond$ Self-reg on $\sN^{(lex)}$ only  & \textbf{39.9} & 91.4 & 38.5 & 90.3 \\
\bottomrule
\end{tabular}
\end{table}

\paragraph{Narrowing Self-regularization Targets.}
By default, we apply the self-reg to hard negatives from both representing modules, which intuitively is a compromise choice for both. To explore if the self-reg can push one of them to an extreme, we conduct exploratory settings for the self-reg in Table~\ref{tab:exp_dive_self_reg}. 
First, applying self-reg to the negatives from dense-vector module even makes the whole framework degenerate. It is likely attributed to the dense-vector receiving less supervisions from the lexicon part, which supports the above claim that the self-reg can be seen as a distillation from lexicons to dense embedding.
On the other hand, when applying self-reg only to the negatives by the lexicon part, the lexicon-based model achieves a new level with 39.9\% MRR@10, which is superior to a single-representing retriever. This supports the idea of instance-dependent prompt learning, where all modules work together for better lexicon-weighting representations.

\begin{figure*}[t]
\begin{center}
\centering
%\subfigure{
\begin{minipage}[t]{0.3\linewidth}
\centering
\includegraphics[width=0.94\linewidth]{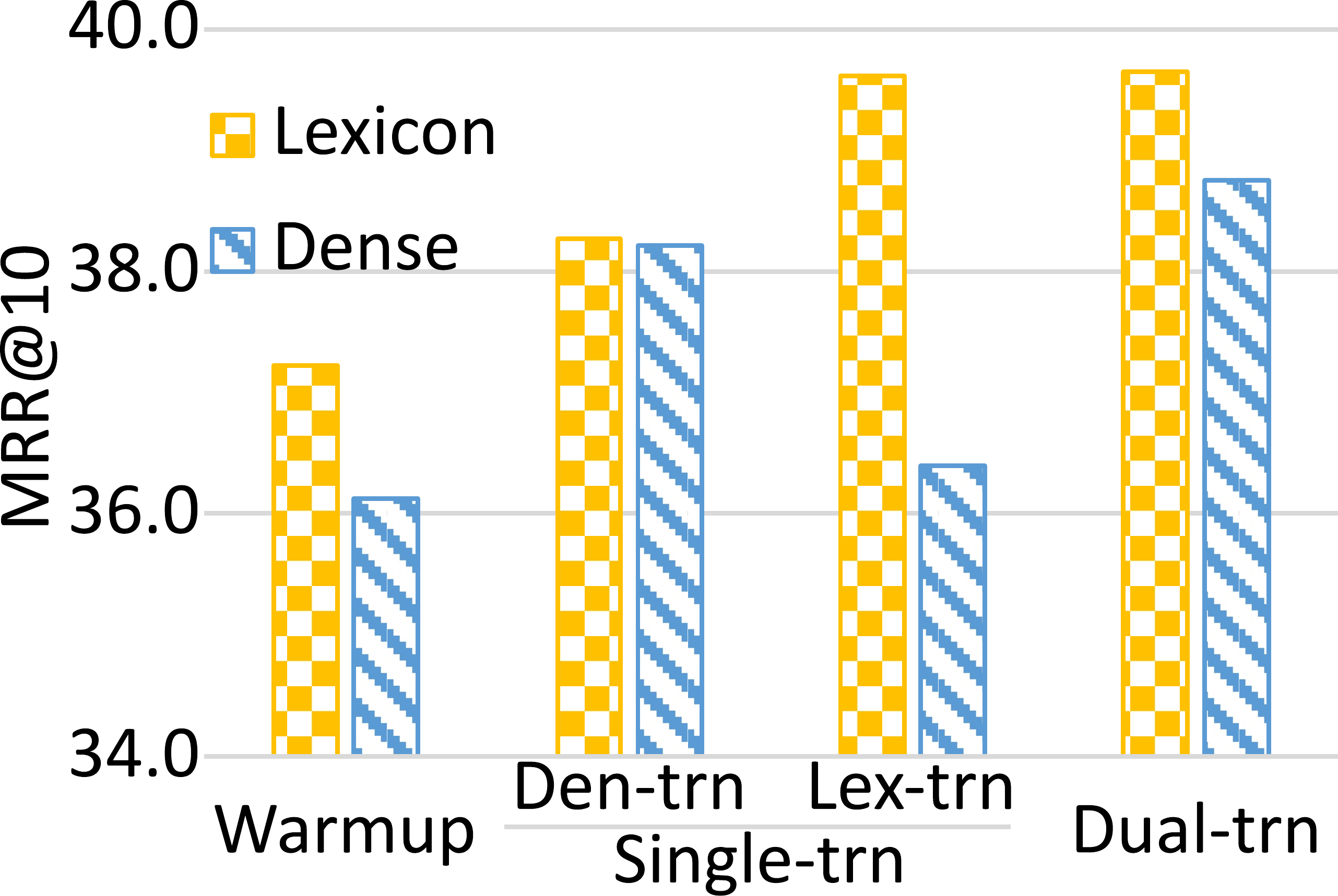}
% \vspace{-8pt}
\caption{\small Verifying consistency of dual representing modules. `trn' denotes `training'.
}\label{fig:exp_consist_training}
\end{minipage}%
%}
\hspace{3mm}
%\subfigure{
\begin{minipage}[t]{0.33\linewidth}
\centering
\includegraphics[width=0.94\linewidth]{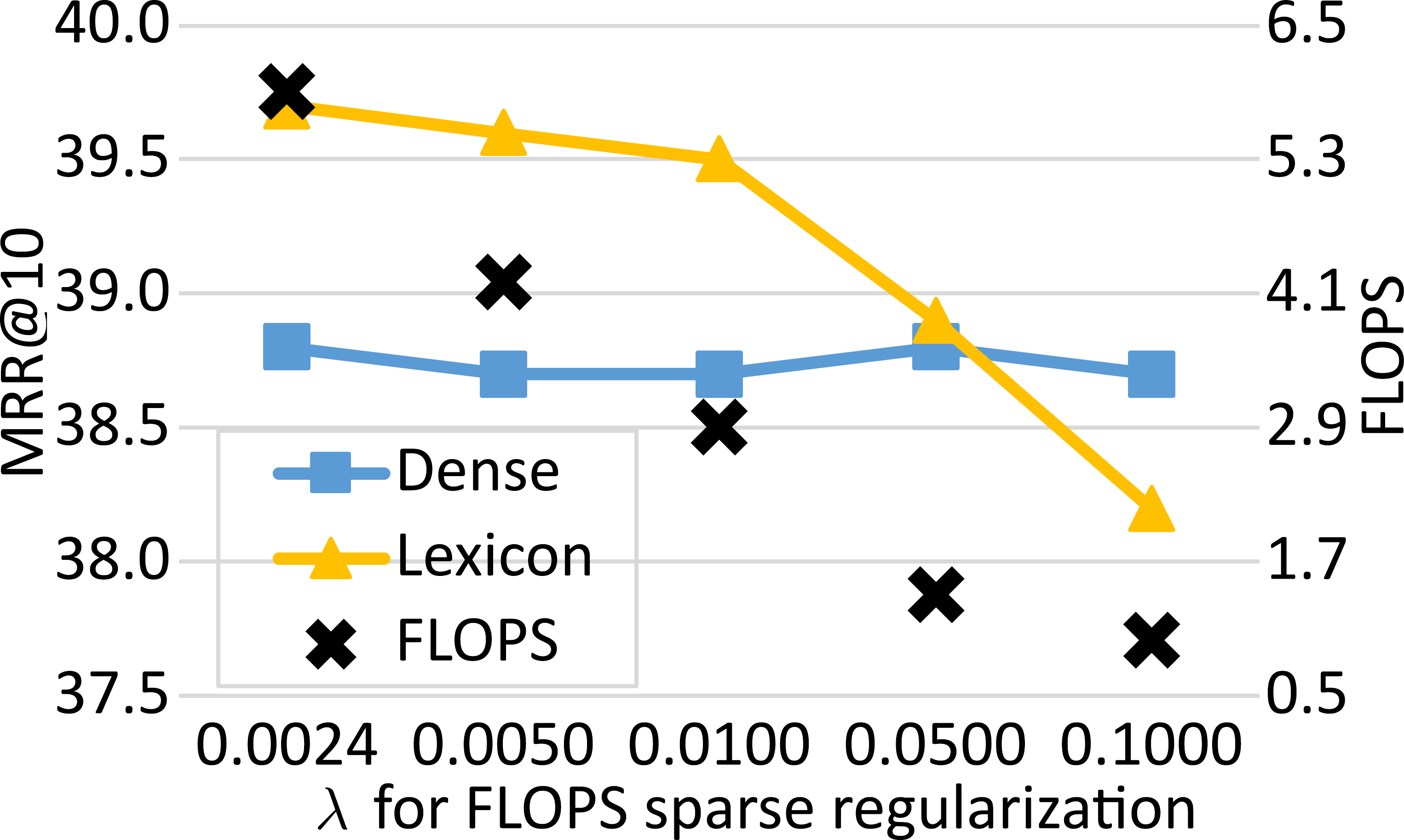}
% \vspace{-8pt}
\caption{\small Effects of the loss weight $\lambda$ of FLOPS sparse regularization on the our performance.}
\label{fig:exp_flops}
\end{minipage}%
%}%
\hspace{3mm}
%\subfigure{
\begin{minipage}[t]{0.30\linewidth}
\centering
\includegraphics[width=0.94\linewidth]{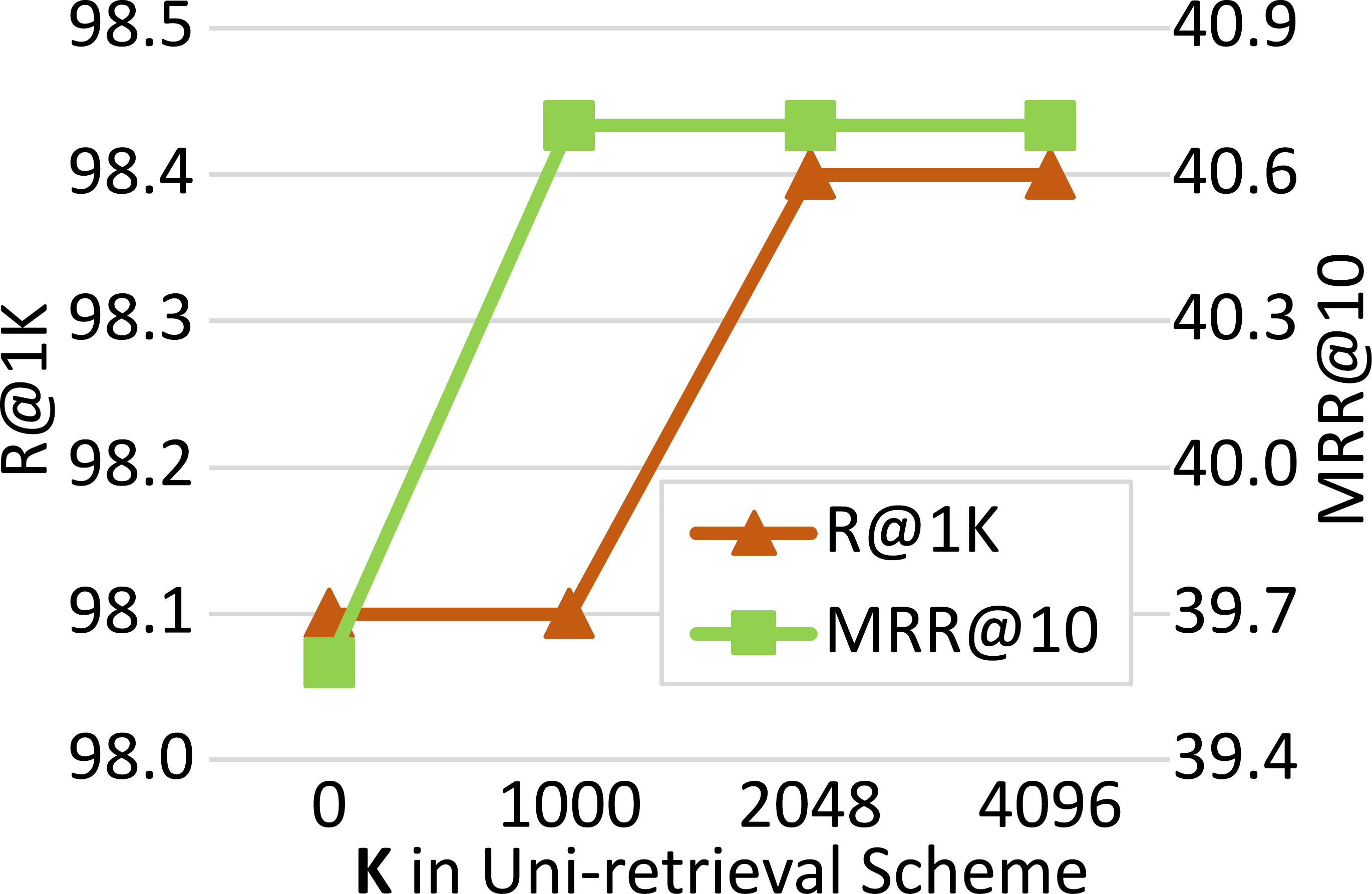}
% \vspace{-8pt}
\caption{\small Effects of the hyperparam K in our uni-retrieval scheme on MS-Marco Dev.}
\label{fig:exp_uni_k}
\end{minipage}%
%}%
\end{center}
% \vspace{-13pt}
%\caption{figure}
% \vspace{-15pt}
% \vspace{-8pt}
\end{figure*}

\paragraph{Evaluation of Learning Consistency. }
To verify if the dual representing modules depend on consistent semantic/syntactic features for the common target, we conduct an experiment to train one of the dual modules but leave the other unchanged at \textit{continual training} stage. 
As in Figure~\ref{fig:exp_consist_training}, the leftmost one is warmup-ed Unife\textsc{r} (warmup), whereas the rightmost one is the full Unife\textsc{r} (dual-trn) as an upper bound of performance. 
Interestingly, optimizing for each of the representing modules can improve both retrieval paradigms (i.e., lexicon and dense). 
This confirms that optimizing one module can benefit the other, attributed to complementary representations and the consistent learning target.

\subsection{Efficiency Analysis}

\paragraph{FLOPS analysis.}
To view sparsity-efficacy trade-off, we vary the loss weight $\lambda$ for FLOPS sparse regularization \cite{Paria2020FLOPS}.
As in Figure~\ref{fig:exp_flops}, with $\lambda$ exponentially increasing, document FLOPS decreases linearly, improving the efficiency of our framework. 
The descending of lexicon-based efficacy is not remarkable when FLOPS > 4 and then becomes notable with the growth of $\lambda$. Fortunately, this will not affect the dense representation in terms of dense-vector retrieval. 

\begin{table}[t]
\small
\centering
\caption{\small Unifie\textsc{R}-lex v.s. QPS by Top-N lexicon sparsifying. The QPS is calculated on a CPU machine with pre-embedded queries, and ORG denotes non-sparsified Unifie\textsc{r}. }
\label{tab:exp_sparsify}
\begin{tabular}{l|ccc|l}
\toprule
\textbf{Top-N} & \textbf{QPS} & \textbf{M@10} & \textbf{R@100} & \textbf{Remark} \\ \hline
BM25	&449	&19.3	&69.0	  & \\\hline
ORG	&50	&41.3	&92.3	& Not sparsified \\\hline
75	&129	&40.8	&91.5	&$\downarrow$ Index as Sparse as BM25 \\
50	&188	&40.4	&91.1	 & \\
25	&343	&38.4	&89.0	& \\
20	&446	&37.5	&87.6	 &$\downarrow$ Infer as Faster than BM25 \\
15	&537	&36.2	&86.0	 & \\
10	&693	&33.6	&82.2	 & \\
8	&911	&31.9	&79.8	 & \\
4	&954	&25.5	&70.0	&$\uparrow$ Better than BM25 \\
2	&1144	&16.2	&53.2	 & \\
1	&1376	&1.8	&22.3	 & \\
\bottomrule
\end{tabular}
% \vspace{-5pt}
% \vspace{-5pt}
\end{table}

\paragraph{Uni-Retrieval Hyperparameter.}
In uni-retrieval scheme, a hyperparam K is used to control computation overheads of dense dot-product. 
As illustrated in Figure~\ref{fig:exp_uni_k}, `K=0' denotes lexicon-only retrieval in Unifie\textsc{r}. 
The table shows that Unifie\textsc{r} reaches an MRR@10 ceiling when K is set to a de facto number, i.e., 1000. 
Then, the upper bound of R@1000 is reached when K=2048. After that, the two metrics cannot be observed with any changes.

\paragraph{Latency Analysis.}

Besides the un-intuitive FLOPS numbers, we also exhibit the latency (measured by `query-per-second' -- QPS) of Unifie\textsc{r}. 
Basically, our Unifie\textsc{r} is bottlenecked by its lexicon head in terms of inference speed as aforementioned, so we would like to dive into the controllable sparsity of Unifie\textsc{r}-lex.
Note that, to reserve a large room for further sparsifying, we leverage the reranker-taught Unifie\textsc{r}-lex as shown in Table~\ref{tab:exp_reranker_distill}, whose MRR@10 is 41.3\%. 
Then, we adopt a simple but effective sparsifying method -- top-N \citep{Yang2021topk} -- but in the index-building process only. As a result, we show the performance of our UnifieR-lexicon with N decreasing in Table~\ref{tab:exp_sparsify}. 
It is shown only the Top-4 tokens kept for each passage can deliver very competitive results with faster speed than BM25.

\subsection{Exploration of Advanced Architecture}
\label{sec:advanced_arc}

\begin{table}[t]
\small
\centering
\caption{\small Stage 1 of Unifie\textsc{r} with query-side gating.}
\label{tab:exp_query_gating}
\begin{tabular}{lcc}
\toprule
{\textbf{Method}} & {\textbf{M@10}} &\textbf{R@100} \\ \hline
Unifie\textsc{r}-uni (warmup) & 38.3 & 90.8 \\
~~+ \textbf{query-side gating} & 39.2 & 91.2 \\ 
\bottomrule
\end{tabular}
% \vspace{-5pt}
% \vspace{-5pt}
\end{table}

\begin{table}[t]
\small
\centering
\setlength{\tabcolsep}{1.5pt}
\caption{\small Reranker-taught Unifie\textsc{r} v.s. previous state-of-the-art (SoTA) models (i.e., Dense \citep{Zhang2021AR2}, Lexicon\cite{Formal2022SPLADE++}, Multi-Vec \cite{Khattab2021ColBERTv2}).}
\label{tab:exp_reranker_distill}
\begin{tabular}{lcccccc}
\toprule
\multirow{2}{*}{{\textbf{Methods}}}  & \multicolumn{2}{c}{Dense-vector} & \multicolumn{2}{c}{Lexicon-based} & \multicolumn{2}{c}{Uni/Multi-Vec} \\
\cmidrule(lr){2-3} \cmidrule(lr){4-5} \cmidrule(lr){6-7}
& \textbf{M@10} &\textbf{R@100} & \textbf{M@10} &\textbf{R@100} & \textbf{M@10} &\textbf{R@100} \\ \hline
Previous SoTA & 39.5 & - & 37.5 & - & 39.7 & - \\
Unifie\textsc{r} & 38.8 & 90.3 & 39.7 & 91.2 & 40.7 & 92.0 \\ \hline
Unifie\textsc{r} (\textbf{distill}) & 40.5 & 91.6 & 41.3 & 92.3 & 42.0 & 93.0 \\ 
\bottomrule
\end{tabular}
% \vspace{-5pt}
% \vspace{-5pt}
\end{table}

\input{tables/case_pos}

\paragraph{Query-side Gating Mechanism. }
As it is too rough to directly add the scores of the two retrieval paradigms, we incorporate a recent inspiration of mix-of-expert (MoE) to enhance the combination of the two paradigms. 
As in illustrated in \S\ref{sec:illu_gate}, we leveraged a gating mechanism to switch Unifie\textsc{r} between dense and lexicon, based solely on the semantics of queries. The reasons for ``solely on queries'' are two-fold: i) the analyses in \S\ref{sec:qalitative_analysis} show that the type of queries affects models a lot 
and ii) the dependency on queries only will not affect the indexing procedure for large-scale collections, leading to zero extra inference overheads. 
After this gating mechanism in the warmup stage of Unifie\textsc{r} training where the gate's optimization is based on the relevance score of uni-retrieval. 
As listed in Table~\ref{tab:exp_query_gating}, a remarkable improvement is observed with such a query-side gating mechanism (+0.9\% MRR@10 and +0.4\% R@100).

\paragraph{Reranker-taught Unifie\textsc{r}.}

Although the Unifie\textsc{r} in Table \ref{tab:exp_main_dev_dl19} \& \ref{tab:exp_main_otherrecall} seems significant in terms of performance improvement, it's noteworthy that the comparisons are unfair because Unifie\textsc{r} didn't use a re-ranker (a strong but heavy cross-encoder) as a teacher for knowledge distillation (see `Reranker taught' in Table~\ref{tab:exp_main_dev_dl19}).
To make the comparisons fairer, we first trained a re-ranker based on Unifie\textsc{r}'s hard negatives and then used a KL loss for distillation in the Continual Training Stage (as illustrated in Figure~\ref{fig:model_v2_pipeline} of \S\ref{sec:rerank_taught_pipeline}). 
As listed in Table~\ref{tab:exp_reranker_distill}, it is shown that i) our proposed Unifie\textsc{r} is compatible with `Reranker taught' scheme and consistently brings 1\%+ improvement, and ii) Unifie\textsc{r} outperforms its strong competitors by large margins (2.0\%+).

\subsection{Qualitative Analysis} \label{sec:qalitative_analysis}
\paragraph{Case Study.} As shown in Table~\ref{tab:case_pos}, we list two queries coupled with the ranking results from five retrieval systems. 
Those are from three groups, i.e., i) previous state-of-the-art dense-vector and lexicon-based retrieval models, ii) the dense-vector and lexicon-based retrieval modules from our Unifie\textsc{r}, and iii) uni-retrieval scheme by our Unifie\textsc{r}.
As demonstrated in the first query of the table, `Indep-lex' achieves a very poor performance, where the positive passage is ranked as 94. Via exhibiting its top-1 passage, the error is possibly caused by the confusion between the `weather' for a specific day and `weather' for a period (i.e., climate). This is because the `weather' as a pivot word in both contexts receives large weights, making the distinguishment very hard.
Although our Unifie\textsc{r}$_\text{lex}$ can lift the positive from 94 to 3 by our carefully designed unified model, it still suffers from confusion. 
Meantime, it is observed that both dense-vector methods perform well since they rely on latent semantic contextualization, less focusing on a specific word. 
As shown in the second query of the table, the strange word, `idiotsguides' makes both dense-vector models less competent. 
On the contrary, the lexicon-based method can handle this case perfectly.
It is still noteworthy that our Unifie\textsc{r}$_\text{den}$ can also outperform the vanilla one, `Indep-den', by lifting 31 (41$\rightarrow$10) ranking position. This is attributed to our consistent feature learning, which bridges the gap of heterogeneity between dense-vector and lexicon-based retrieval. 
These two cases also support the previous claim that the two representing ways can provide distinct views of query-document relevance. 
Furthermore, despite varying performance across different paradigms, our uni-retrieval scheme consistently performs well as it is an aggregation of both. 
Please see \S\ref{sec:error_ana} for our further case study on error analysis.

\paragraph{Limitation.}

The main limitations of this work are i) \textit{PLM Compatibility}: due to the special encoder design, Unifie\textsc{r} can only be initialized from a limited number of pre-trained models, 
and ii) \textit{Additional Infrastructure}: in spite of the almost same retriever latency as traditional lexicon-based retrieval, Unifie\textsc{r} requires extra computation infrastructure for indexing and storing both dense and sparse embeddings of all documents in the collection.

\section{Conclusion} \label{sec:conclusion}

We present a brand-new learning framework, dubbed Unifie\textsc{r}, to unify dense-vector and lexicon-based representing paradigms for large-scale retrieval.
It improves the two paradigms by a carefully designed neural encoder to fully exploit the representing capability of pre-trained language models.
Its capability is further strengthened by our proposed dual-consistency learning with self-adversarial and -regularization. 
Moreover, the uni-retrieval scheme and the advanced architectures upon our encoder are presented to achieve more.  
Experiments on several benchmarks verify the effectiveness and versatility of our framework.

\bibliographystyle{ACM-Reference-Format}
\bibliography{reference}

\appendix

\input{tables/beir}

\section{Explanation of Two Recall Metrics} \label{sec:recall_metric}
Regarding R@N metric, we found there are two kinds of calculating ways, and we strictly follow the official evaluation one at
\url{https://github.com/usnistgov/trec_eval} and 
\url{https://github.com/castorini/anserini}, which is defined as 
\begin{align}
    \text{Marco-Recall@N} = \dfrac{1}{|\sQ|} \sum_{q\in\sQ} \dfrac{ \sum_{d_+\in \sD_+} \1_{d_+ \in \bar\sD} }{\min(N, |\sD_+|)},
\end{align}
where there may be multiple positive documents $\sD^+\in\sD$, $\sQ$ denotes the test queries and $\bar\sD$ denotes top-K document candidates by a retrieval system. We also call this metric \textit{all-positive-macro} Recall@N. 
On the other hand, another recall calculation method following DPR \citep{Karpukhin2020DPR} is defined as 
\begin{align}
    \text{DPR-Recall@N} = \dfrac{1}{|\gQ|} \sum_{q\in\sQ} \1_{\exists d \in \bar\sD \wedge d \in \sD^+ }.
\end{align}
which we call \textit{one-positive-enough} Recall@N. 
Therefore, The official (\textit{all-positive-macro}) Recall@N is usually less than DPR (\textit{one-positive-enough}) Recall@N, and the smaller N, the more obvious.

\section{Supplementary Experiment Supports}

\subsection{BEIR Details} \label{sec:beir_full}

Please refer to Table~\ref{tab:exp_beir_details} for detailed results on BEIR benchmark with 12 datasets. 
It is noteworthy that applying a retriever trained on legacy data (e.g., MS-MARCO labeled in 2016) to the latest topics (e.g., TREC-COVID after 2020) will likely be vulnerable to distribution shifts over time. 
Because rare or/and brand-new words are usually present over time, we argue that the proposed UnifieR would suffer less from the out-of-vocab (OOV) problem during the distribution shifts. Basically, as we adopted the BERT tokenizer with WordPiece techniques \citep{Song2021wordpiece}, our neural retriever can still model the input text without any interference: Consistent with BERT \citep{Devlin2019BERT} and coCondender \citep{Gao2021coCondenser} pre-training, some rare words will be split into smaller units (e.g., word "idiotsguide" split into `idiots' and `\#\#guide') before being fed into the Transformer architecture, so obvious OOV problem will not arise. In addition, many previous works \citep{Formal2021SPLADE,Formal2021SPLADEv2,Formal2022SPLADE++,Hofstatter2020Margin-MSE,Thakur2021BEIR,Zhou2022Hyperlink} found that lexicon-based methods can achieve better zero-shot retrieval performance because ``lexicon overlap'' is an essential feature for them to calculate query-document relevance scores, aligning closely with the goal of first-stage retrieval. Therefore, inheriting the advantages from previous lexicon-based methods, our lexicon retrieval head could alleviate the distribution shifts over time.

% \section{More Explorations}

\subsection{Illustration of Query-side Gating} \label{sec:illu_gate}

\begin{figure}[t]
    \centering
    \includegraphics[width=0.85\linewidth]{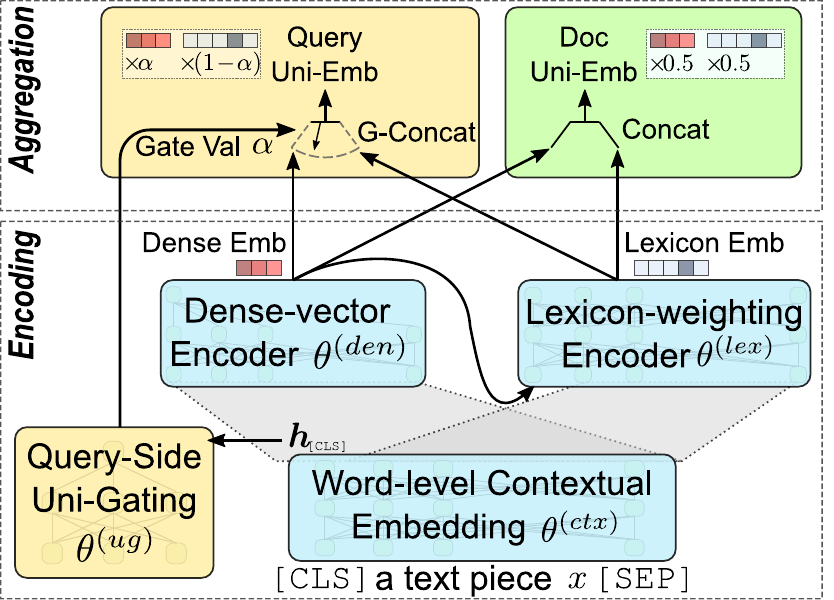}
    \caption{\small Equipping Unifie\textsc{r} with query-side gating. }
    \label{fig:model_v2_arc}
\end{figure}

We illustrate the query-side gating mechanism in Figure~\ref{fig:model_v2_arc}, which leverages a gating mechanism to dynamically combine lexicon and dense embeddings only at the query side.

\subsection{Reranker-taught Pipeline} \label{sec:rerank_taught_pipeline}

\begin{figure}[t]
    \centering
    \includegraphics[width=1.0\linewidth]{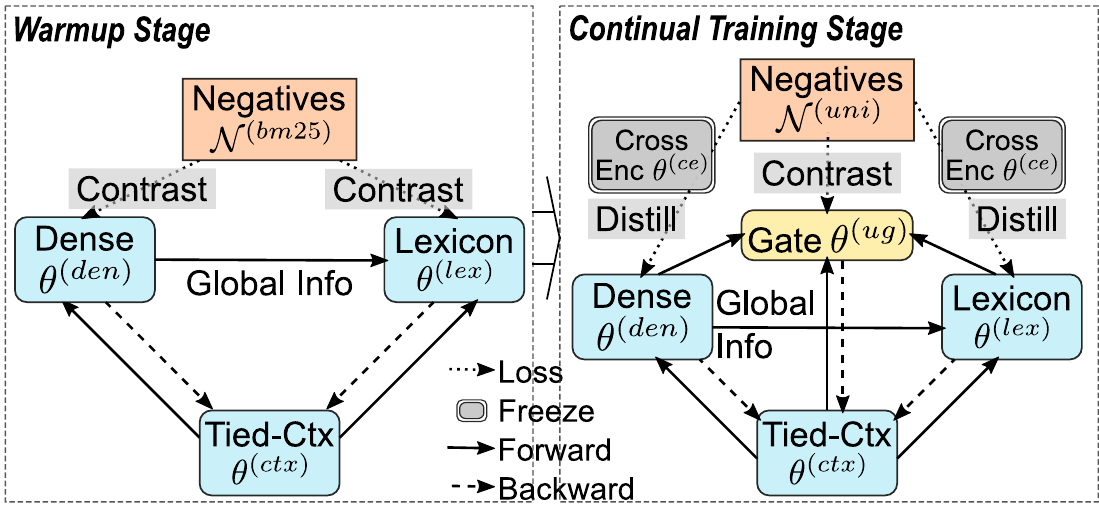}
    \caption{\small Reranker-taught Unifie\textsc{r} by knowledge distillation. }
    \label{fig:model_v2_pipeline}
\end{figure}

In contrast to the normal two-stage training pipeline in Figure~\ref{fig:model_learning}, we present our reranker-taught pipeline in Figure~\ref{fig:model_v2_pipeline}. 

\input{tables/case_errors}

\subsection{Error Cases} \label{sec:error_ana}

As shown in Table~\ref{tab:case_errors}, we show two representative cases which our proposed method cannot handle. 

i) \textit{query hubness}: The first case shows a query that cannot be tackled by our Unifie\textsc{r} in any retrieval paradigm. However, it is observed that the top-1 passage retrieved by our model can also be considered as a positive passage, which can answer the query `\textit{what is a dvt}'. 
These negative passages for the query are false negatives, which are brought by the limited crowd-sourcing labeling procedure. 
Therefore, the poor performance of our model instead proves that our model is more robust, whereas the independent learning model is overfitting to its false negatives, resulting in seemingly good outputs.

ii) \textit{Insufficient representation ability}: The second case lists the top-retrieved passages for all five retrieval systems. 
It is shown that compared to independently learned retrieval models (i.e., `Indep-den' and `Indep-lex'), our unified models even perform worse and retrieve less relevant passages (refer to Unifie\textsc{r}$_\text{den}$'s 1st). 
An interesting point is that the `Ups'-related passage is retrieved by our Unifie\textsc{r}$_\text{den}$ since `upsell' is tokenized as `ups' and `\#\#ell'. 
This is highly likely since one single model is required to serve dual representing modules, compromising its representation ability. 

Meantime, our uni-retrieval can still improve the ranking performance by combining both of the representing worlds.

\end{document}

%% file: math_commands.tex
\usepackage{amsmath,amsfonts,bm}

\def\eqref#1{equation~\ref{#1}}
\def\floor#1{\lfloor #1 \rfloor}
\def\1{\bm{1}}

\def\rd{{\textnormal{d}}}

\def\vh{{\bm{h}}}

\def\vp{{\bm{p}}}

\def\vu{{\bm{u}}}
\def\vv{{\bm{v}}}

\def\mH{{\bm{H}}}

\def\mW{{\bm{W}}}

\DeclareMathAlphabet{\mathsfit}{\encodingdefault}{\sfdefault}{m}{sl}
\SetMathAlphabet{\mathsfit}{bold}{\encodingdefault}{\sfdefault}{bx}{n}

\def\gQ{{\mathcal{Q}}}

\def\sD{{\mathbb{D}}}
\def\sN{{\mathbb{N}}}

\def\sQ{{\mathbb{Q}}}

\def\sV{{\mathbb{V}}}

\newcommand{\R}{\mathbb{R}}

\newcommand{\KL}{D_{\mathrm{KL}}}

%% file: define-operator.tex
\usepackage{mathtools}
\usepackage{multirow}
\usepackage{multicol}
\usepackage{booktabs}
\usepackage{subfigure}

\DeclareMathOperator*{\enc}{Enc}

\DeclareMathOperator*{\denseenc}{Uni-Den}
\DeclareMathOperator*{\lexiconenc}{Uni-Lex}

\DeclareMathOperator*{\transformerenc}{Transfm-Enc}

\DeclareMathOperator*{\pool}{Pool}

\DeclareMathOperator*{\maxpool}{Max-Pool}

\DeclareMathOperator*{\relu}{ReLU}

%% file: tables/main_dev_update1.tex
% \usepackage{multirow}
% \usepackage[normalem]{ulem}
\begin{table*}[t]
\centering
\small
% \setlength{\tabcolsep}{1.3pt}
% \vspace{-6pt}
\caption{\small Passage retrieval results on MS-Marco Dev and TREC Deep Learning 2019. 
\dag Refer to Table~\ref{tab:exp_main_otherrecall}. 
`coCon': coCondenser that continually pre-trained BERT in unsupervised manner. 
`Reranker taught': distillation from a reranker (see \S\ref{sec:rel_work}).
}\label{tab:exp_main_dev_dl19}
\begin{tabular}{llcccccccc}
\toprule
\multicolumn{1}{c}{\multirow{2}{*}{\textbf{Method}}} & \multicolumn{1}{c}{\multirow{2}{*}{\textbf{\makecell{Pre-trained \\ model}}}} & \multirow{2}{*}{\textbf{\makecell{Reranker \\   taught}}} &  \multirow{2}{*}{\textbf{\makecell{Hard \\ negs}}} & \multirow{2}{*}{\textbf{\makecell{Mul \\ Repr}}} & \multicolumn{3}{c}{\textbf{MS-Marco Dev}}     & \multicolumn{2}{c}{\textbf{TREC DL 19}} \\ 
% \cline{5-7} \cline{8-9}
\cmidrule(lr){6-8} \cmidrule(l){9-10}
\multicolumn{1}{c}{}                                  & \multicolumn{1}{c}{}                              &                                                            &                                                &   & MRR@10        & R@100         & R@1k          & R@100              & nDCG@10            \\ \hline \hline
\multicolumn{9}{l}{\textit{Dense-vector   Retriever}}                                                                                                                                                                                                                                                                \\ \hline
ANCE~\cite{Xiong2021ANCE}                             & RoBERTa$_\text{base}$                             &                                                            &                                                &   & 33.8          & 86.2          & 96.0          & 44.5               & 65.4               \\
% LTRe~\cite{Zhan2020LTRe}                              & RoBERTa$_\text{base}$                             &                                                            &                                               &    & 32.9          & -             & 95.5          & -                  & 66.1               \\
% ANCE + LTRe~\cite{Zhan2020LTRe}                       & RoBERTa$_\text{base}$                             &                                                            &                                               &    & 34.1          & -             & 96.2          & -                  & 67.5               \\
% Margin-MSE~\cite{Hofstatter2020Margin-MSE}            & DistilBERT                                        & $\checkmark$                                               &                                                &   & 32.3          & -             & 95.7          & -                  & 69.7               \\
ADORE~\cite{Zhan2021STAR-ADORE}                       & RoBERTa$_\text{base}$                             &                                                            & $\checkmark$                            &          & 34.7          & 87.6          & -             & 47.3               & 68.3               \\
TAS-B~\cite{Hofstatter2021TAS-B}                      & DistilBERT                                        & $\checkmark$                                               &                                                &   & 34.7          & -             & 97.8          & -                  & 71.2               \\
TCT~\cite{Lin2021TCT}                                 & BERT$_\text{base}$                                & $\checkmark$                                               &                                                   && 33.5          & -             & 96.4          & -                  & 67.0               \\
TCT-ColBERT~\cite{Lin2021TCT}                         & BERT$_\text{base}$                                & $\checkmark$                                               & $\checkmark$                                &      & 35.9          & -             & 97.0          & -                  & 71.9               \\
Condenser~\cite{Gao2021Condenser}                     & Condenser$_\text{base}$                                &                                                            & $\checkmark$                                     & & 36.6          & -             & 97.4          & -                  & -                  \\
% Condenser-stage1~\cite{Gao2021Condenser}              & BERT$_\text{base}$                                &                                                            &                                                   & 33.8          & -             & 96.1          & -                  & -                  \\
coCondenser~\cite{Gao2021coCondenser}                 & coCon$_\text{base}$                                &                                                            & $\checkmark$                                &      & 38.2          & -             & \textbf{98.4} & -                  & -                  \\
% coCondenser-stage1~\cite{Gao2021coCondenser}          & BERT$_\text{base}$                                &                                                            &                                                   & 35.7          & -             & 97.8          & -                  & -                  \\
% coCondenser (ours)                                    & BERT$_\text{base}$                                &                                                            & $\checkmark$                                      &               &               &               &                    &                    \\
ColBERTv1~\cite{Khattab2020ColBERT}                   & BERT$_\text{base}$                                &                                                            &                                                 & $\checkmark$  & 36.0          & -             & 96.8          & -                  & -                  \\
ColBERTv2~\cite{Khattab2021ColBERTv2}                 & BERT$_\text{base}$                                & $\checkmark$                                               & $\checkmark$                                  & $\checkmark$   & {\ul 39.7}    & -             & \textbf{98.4} & -                  & -                  \\
PAIR~\cite{Ren2021PAIR}                               & ERNIE$_\text{base}$                               & $\checkmark$                                               &                                               &    & 37.9          & -             & -\dag             & -                  & -                  \\
RocketQA~\cite{Qu2021RocketQA}                        & ERNIE$_\text{base}$                               & $\checkmark$                                               &                                               &    & 37.0          & -             & -\dag            & -                  & -                  \\
RocketQAv2~\cite{Ren2021RocketQAv2}                   & ERNIE$_\text{base}$                               & $\checkmark$                                               & $\checkmark$                                &      & 38.8          & -             & -\dag             & -                  & -                  \\
AR2~\cite{Zhang2021AR2}                               & coCon$_\text{base}$                                       & $\checkmark$                                               & $\checkmark$                                 &     & 39.5          & -             & -\dag            & -                  & -                  \\ \hline \hline
\multicolumn{9}{l}{\textit{Lexicon-base or Sparse Retriever}}                                                                                                                                                                                                                                                                \\ \hline
DeepCT~\cite{Dai2019DeepCT}                           & BERT$_\text{base}$                                &                                                            &                                              &     & 24.3          & -             & 91.3          & -                  & 55.1               \\
RepCONC~\cite{Zhan2022RepCONC}                        & RoBERTa$_\text{base}$                             &                                                            & $\checkmark$                                &      & 34.0          & 86.4          & -             & 49.2               & 66.8               \\
SPLADE-max~\cite{Formal2021SPLADEv2}                  & DistilBERT                                        &                                                            &                                              &     & 34.0          & -             & 96.5          & -                  & 68.4               \\
SPLADE-doc~\cite{Formal2021SPLADEv2}                  & DistilBERT                                        &                                                            &                                           &        & 32.2          & -             & 94.6          & -                  & 66.7               \\
DistilSPLADE-max~\cite{Formal2021SPLADEv2}            & DistilBERT                                        & $\checkmark$                                               &                                              &     & 36.8          & -             & 97.9          & -                  & 72.9               \\
SelfDistil~\cite{Formal2022SPLADE++}                  & DistilBERT                                        & $\checkmark$                                               & $\checkmark$                      &                & 36.8          & -             & 98.0          & -                  & 72.3               \\
EnsembleDistil~\cite{Formal2022SPLADE++}              & DistilBERT                                        & $\checkmark$                                               & $\checkmark$                      &                & 36.9          & -             & 97.9          & -                  & 72.1               \\
Co-SelfDistil~\cite{Formal2022SPLADE++}               & coCon$_\text{base}$                                       & $\checkmark$                                               & $\checkmark$                       &               & 37.5          & -             & \textbf{98.4} & -                  & 73.0               \\
Co-EnsembleDistil~\cite{Formal2022SPLADE++}           & coCon$_\text{base}$                                       & $\checkmark$                                               & $\checkmark$                          &    $\checkmark$        & 38.0          & -             & {\ul 98.2}    & -                  & 73.2               \\ \hline \hline
\multicolumn{9}{l}{\textit{Hybrid Retriever}}                                                                                                                                                                                                                                                                        \\ \hline
CLEAR~\cite{Gao2021CLEAR}                             & BERT$_\text{base}$                                &                                                            &                                               & $\checkmark$   & 33.8          & -             & 96.9          & -                  & 69.9               \\
COIL-full~\cite{Gao2021COIL}                          & BERT$_\text{base}$                                &                                                            &                                              &  $\checkmark$   & 35.5          & -             & 96.3          & -                  & 70.4               \\ \hline \hline

Unifie\textsc{r}$_{\text{lexicon}~~(\textit{warmup})}$        & coCon$_\text{base}$                                       &                                                            &                               &       & 37.2    & 90.1    & 97.8          & 50.1         & 69.7        \\
Unifie\textsc{r}$_{\text{dense}~~(\textit{warmup})}$                 & coCon$_\text{base}$                                       &                                                            &                          &           & 36.1          & 87.7          &96.6          & 44.6              & 63.9               \\ 
Unifie\textsc{r}$_{\text{uni-retrieval}~~(\textit{warmup})}$       & coCon$_\text{base}$                                       &                                                            &                                &   $\checkmark$  & 38.3 & 90.8 & 98.0 & 50.6      & 70.2      \\ \hline

Unifie\textsc{r}$_{\text{lexicon}}$                   & coCon$_\text{base}$                                       &                                                            & $\checkmark$                                   &   & {\ul 39.7}    & {\ul 91.2}    & 98.1          & {\ul 53.2}         & {\ul 73.3}         \\
Unifie\textsc{r}$_{\text{dense}}$                     & coCon$_\text{base}$                                       &                                                            & $\checkmark$                                  &    & 38.8          & 90.3          & 97.6          & 50.2               & 71.1               \\ 
Unifie\textsc{r}$_{\text{uni-retrieval}}$             & coCon$_\text{base}$                                       &                                                            & $\checkmark$                                  &  $\checkmark$  & \textbf{40.7} & \textbf{92.0} & \textbf{98.4} & \textbf{53.8}      & \textbf{73.8}      \\
\bottomrule
\end{tabular}
% \vspace{-6pt}
% \vspace{-8pt}
\end{table*}

%% file: tables/ablation.tex
\begin{table}[t]\small
\centering
\caption{\small Ablation of the encoder on MS-Marco Dev.}
\label{tab:exp_abla_encoder}
\begin{tabular}{lcccc}
\toprule
\multirow{2}{*}{{\textbf{Methods}}} &  \multicolumn{2}{c}{{Lexicon-based}} & \multicolumn{2}{c}{{Dense-vector}} \\
% \cline{2-3} \cline{4-5}
\cmidrule(lr){2-3} \cmidrule(l){4-5}
& {M@10} & {R@100}  & {M@10} & {R@100} \\
\hline
Unifie\textsc{r} \textit{(warmup)} & 37.2 & 90.1 & 36.1 & 87.7 \\
~$\Diamond$ w/o sharing Global & 36.1 & 89.8 & 35.2 & 87.2 \\
~$\Diamond$ w/o in-depth Interact & 36.1 & 89.3 & 35.7 & 89.7 \\
\bottomrule
\end{tabular}
% \vspace{-5pt}
% \vspace{-5pt}
\end{table}

%% file: tables/case_pos.tex
\begin{table*}[t]
\centering
\small
\setlength{\tabcolsep}{1pt}
\caption{\small Case study on MS-Marco Dev set. 
`Passage+' denotes positive passage of the corresponding query. 
`Indep-den' denotes a well-trained state-of-the-art dense-vector retrieval model with static hard negatives (i.e., coCondenser \cite{Gao2021coCondenser}, M@10=38.2) while `Indep-lex' denotes a well-trained state-of-the-art lexicon-based retrieval model with static hard negatives (i.e., SPLADE \cite{Formal2022SPLADE++}, M@10=38.5).
}\label{tab:case_pos}
\begin{tabular}{lp{16.2cm}}
\toprule
\textbf{Query} & \textit{ID:1088347//} weather in new york city ny \\\hline
\textbf{Passage+} & 
\textit{ID:7094280//} \textit{Title:} - \textit{Body:} New York, NY - Weather forecast from Theweather.com. Weather conditions with updates on temperature, humidity, wind speed, snow, pressure, etc. for New York, New York Today: Cloudy skies with light rain, with a maximum temperature of 72C and a minimum temperature of 52C.\\ \hline
\textbf{Rank} & 
Indep-den: ~1;~~~
Indep-lex: 94;~~~
Unifie\textsc{r}$_\text{den}$: ~1;~~~ Unifie\textsc{r}$_\text{lex}$: ~3;~~~ Unifie\textsc{r}$_\text{uni}$: ~1~~~\\ \hline
\textbf{Retrieved} & 
\textbf{Indep-lex's 1st.} \textit{ID:65839//} \textit{Title:} New York City - Best Time To Go \& When to Go \textit{Body:} Weather: Spring in New York City is the best time to be in the city, without doubt. Spring usually means less humidity and temps between 50-80 degrees, though June occasionally sees a 90 degree day. An occasional humidity soaked heat wave can strike, but it usually feels nice the first time around. \\
& \textbf{Indep-lex's 2nd.} \textit{ID:4835773//} \textit{Title:} Climate of New York \textit{Body:} Weather: Unlike the vast majority of the state, New York City features a humid subtropical climate (Koppen Cfa). New York City is an urban heat island, with temperatures 5-7 degrees Fahrenheit (3-4 degrees Celsius) warmer overnight than surrounding areas. In an effort to fight this warming, roofs of buildings are being painted white across the city in an effort to increase the reflection of solar energy, or albedo. \\
& \textbf{Unifie\textsc{r}$_\text{lex}$'s 1st.} \textit{ID:65839//} \textit{Title:} New York City - Best Time To Go \& When to Go \textit{Body:} Weather: Spring in New York City is the best time to be in the city, without doubt. Spring usually means less humidity and temps between 50-80 degrees, though June occasionally sees a 90 degree day. An occasional humidity soaked heat wave can strike, but it usually feels nice the first time around. \\
& \textbf{Unifie\textsc{r}$_\text{lex}$'s 2nd.} \textit{ID:8819213//} \textit{Title:} New York City - Best Time To Go \& When to Go \textit{Body:} Weather: Spring in New York City is the best time to be in the city, without doubt. Spring usually means less humidity and temps between 50-80 degrees, though June occasionally sees a 90 degree day.  \\
% \hline
\toprule
\textbf{Query} & \textit{ID:391101//} idiotsguides tai chi \\\hline
\textbf{Passage+} & 
\textit{ID:7668258//} \textit{Title:} - \textit{Body:} Bill is the author of The Complete Idiot's Guide to T'ai Chi \& Qigong (4th edition), and his newest upcoming books, The Tao of Tai Chi, and The Gospel of Science, in which he paints a vision of vast global benefit as mind-body sciences spread across the planet. \\ \hline
\textbf{Rank} & 
Indep-den: 41;~~~
Indep-lex: ~1;~~~
Unifie\textsc{r}$_\text{den}$: 10;~~~ 
Unifie\textsc{r}$_\text{lex}$: ~1;~~~ 
Unifie\textsc{r}$_\text{uni}$: ~1~~~\\ \hline
\textbf{Retrieved} 
& \textbf{Indep-den's 1st.} \textit{ID:1603205//} \textit{Title:} - \textit{Body:} Tai chi. Tai chi (simplified Chinese: ; traditional Chinese: ; pinyin: chi, an abbreviation of ;is an internal Chinese martial art (Chinese: ; pinyin: ) practiced for both its defense training and its health benefits.  \\ 
& \textbf{Indep-den's 2nd.} \textit{ID:3449438//} \textit{Title:} Tai chi: A gentle way to fight stress \textit{Body:} Tai chi is an ancient Chinese tradition that, today, is practiced as a graceful form of exercise. It involves a series of movements performed in a slow, focused manner and accompanied by deep breathing. Tai chi, also called tai chi chuan, is a noncompetitive, self-paced system of gentle physical exercise and stretching. \\ 
& \textbf{Unifie\textsc{r}$_\text{den}$'s 1st.} \textit{ID:2294942//} \textit{Title:} WHAT IS TAI CHI? \textit{Body:} The Chinese characters for Tai Chi Chuan can be translated as the `Supreme Ultimate Force'. The notion of `supreme ultimate' is often associated with the Chinese concept of yin-yang, the notion that one can see a dynamic duality (male/female, active/passive, dark/light, forceful/yielding, etc.) in all things. \\ 
& \textbf{Unifie\textsc{r}$_\text{den}$'s 2nd.} \textit{ID:3449442//} \textit{Title:} What is Tai Chi? \textit{Body:} What is Tai Chi? In China, and increasingly throughout the rest of the world, tai chi is recognized for its power to instill and maintain good health and fitness in people of all ages. Tai chi aims to bring balance to body, mind and spirit through specifically designed movements, natural breathing and a calm state of mind. It is easily recognized by its slow, captivating and mesmerizing movements. It represents a way of life, helping people meet day to day challenges while remaining calm and relaxed. \\ 
\bottomrule
\end{tabular}
\end{table*}

%% file: tables/beir.tex
\begin{table*}[t]
\small 
\centering
\caption{\small Detailed results (NDCG@10) on BEIR benchmark. 
% [\texttt{UC}] is UniCOIL, [\texttt{CB}] is ColBERT, [\texttt{TB}] is TAS-B, [\texttt{CT}] is Contriever
} \label{tab:exp_beir_details}
\begin{tabular}{lcccccccccc}
\toprule
\multicolumn{1}{c}{\multirow{2}{*}{\textbf{Methods}}} & \multicolumn{4}{c}{\textbf{Sparse}}                  & \multicolumn{6}{c}{\textbf{Dense}}                                     \\ \cmidrule(l){2-5} \cmidrule(l){6-10}
\multicolumn{1}{c}{}                                  & BM25          & DT5Q & UniCOIL & ColBERT       & DPR   & ANCE  & GenQ          & TAS-B & Contriever   & Ours           \\ \midrule
% In-Domain & & & & & & & & & & & \\
% \hline
TREC-COVID                                            & 65.6          & 71.3       & 59.7    & 67.7          & 33.2  & 65.4  & 61.9          & 48.1  & 59.6          & \textbf{71.5}  \\
NFCorpus                                              & 32.5          & 32.8       & 32.5    & 30.5          & 18.9  & 23.7  & 31.9          & 31.9  & 32.8          & \textbf{32.9}  \\
NQ                                                    & 32.9          & 39.9       & 36.2    & \textbf{52.4} & 47.4  & 44.6  & 35.8          & 46.3  & 49.8          & 51.4           \\
HotpotQA                                              & 60.3          & 58.0       & 64.0    & 59.3          & 39.1  & 45.6  & 53.4          & 58.4  & 63.8          & \textbf{66.1}  \\
FiQA                                                  & 23.6          & 29.1       & 27.0    & 31.7          & 11.2  & 29.5  & 30.8          & 30.0  & \textbf{32.9} & 31.1           \\
ArguAna                                               & 31.5          & 34.9       & 35.5    & 23.3          & 17.5  & 41.5  & \textbf{49.3} & 42.9  & 44.6          & 39.0           \\
Tóuche-2020                                           & \textbf{36.7} & 34.7       & 25.9    & 20.2          & 13.1  & 24.0  & 18.2          & 16.2  & 23.0          & 30.2           \\
DBPedia                                               & 31.3          & 33.1       & 30.2    & 39.2          & 26.3  & 28.1  & 32.8          & 38.4  & \textbf{41.3} & 40.6           \\
Scidocs                                               & 15.8          & 16.2       & 13.9    & 14.5          & 7.7   & 12.2  & 14.3          & 14.9  & \textbf{16.5} & 15.0           \\
Fever                                                 & 75.3          & 71.4       & 72.3    & \textbf{77.1} & 56.2  & 66.9  & 66.9          & 70.0  & 75.8          & 69.6           \\
Climate-FEVER                                         & 21.3          & 20.1       & 15.0    & 18.4          & 14.8  & 19.8  & 17.5          & 22.8  & \textbf{23.7} & 17.5           \\
SciFact                                               & 66.5          & 67.5       & 67.4    & 67.1          & 31.8  & 50.7  & 64.4          & 64.3  & 67.7          & \textbf{68.6}  \\ \midrule
BEST ON                                               & 1             & 0          & 0       & 2             & 0     & 0     & 1             & 0     & 4             & 4              \\
AVERAGE                                               & 41.1         & 42.4      & 40.0     & 41.8         & 26.4 & 37.7 & 39.8         & 40.4 & 44.3         & \textbf{44.5} \\ \bottomrule
\end{tabular}
\end{table*}

%% file: tables/case_errors.tex
\begin{table*}[ht]
\centering
\small
\setlength{\tabcolsep}{2pt}
\caption{Error analysis on MS-Marco Dev set. 
}\label{tab:case_errors}
\begin{tabular}{lp{15.5cm}}
\toprule
\textbf{Query} & \textit{ID:682365//} what is a dvt? \\\hline
\textbf{Passage+} & 
\textit{ID:7544458//} \textit{Title:} Deep vein thrombosis \textit{Body:} For other uses, see DVT (disambiguation). Deep vein thrombosis, or deep venous thrombosis (DVT), is the formation of a blood clot (thrombus) within a deep vein, most commonly the legs. Nonspecific signs may include pain, swelling, redness, warmness, and engorged superficial veins. \\ \hline
\textbf{Rank} & 
Indep-den: ~3;~~~
Indep-lex: ~2;~~~
Unifie\textsc{r}$_\text{den}$: 12;~~~ 
Unifie\textsc{r}$_\text{lex}$: 11;~~~ 
Unifie\textsc{r}$_\text{uni}$: ~9~~~\\ \hline
\textbf{Retrieved}
& \textbf{Unifie\textsc{r}$_\text{den}$'s 1st.} \textit{ID:5404002//} \textit{Title:} Definition of `DVT' \textit{Body:} Definition of `DVT'. DVT is a serious medical condition caused by blood clots in the legs moving up to the lungs. DVT is an abbreviation for 'deep vein thrombosis'. The results from one of the largest studies yet carried out leave little doubt that DVT is caused by flying. \\ 
& \textbf{Unifie\textsc{r}$_\text{lex}$'s 1st.} \textit{ID:8492523//} \textit{Title:} What Is DVT? \textit{Body:} What Is DVT? Deep vein thrombosis is a blood clot that forms inside a vein, usually deep within your leg. About half a million Americans every year get one, and up to 100,000 die because of it. The danger is that part of the clot can break off and travel through your bloodstream.  \\ 
& \textbf{Unifie\textsc{r}$_\text{uni}$'s 1st.} \textit{ID:8492523//} \textit{Title:} What Is DVT? \textit{Body:} What Is DVT? Deep vein thrombosis is a blood clot that forms inside a vein, usually deep within your leg. About half a million Americans every year get one, and up to 100,000 die because of it. The danger is that part of the clot can break off and travel through your bloodstream.  \\ 
\toprule
\textbf{Query} & \textit{ID:1029124//} what is upsell \\\hline
\textbf{Passage+} & 
\textit{ID:7220016//} \textit{Title:} Upselling \textit{Body:} What is Upselling? Upselling is a sales technique aimed at persuading customers to purchase a more expensive, upgraded or premium version of the chosen item or other add-ons for the purpose of making a larger sale. eCommerce businesses often combine upselling and cross-selling techniques in an attempt to increase order value and maximize profit. \\ \hline
\textbf{Rank} & 
Indep-den: ~3;~~~
Indep-lex: ~1;~~~
Unifie\textsc{r}$_\text{den}$: 11;~~~ 
Unifie\textsc{r}$_\text{lex}$: ~9;~~~ 
Unifie\textsc{r}$_\text{uni}$: ~8~~~\\ \hline
\textbf{Retrieved}
& \textbf{Indep-den's 1st.} \textit{ID:6288350//} \textit{Title:} - \textit{Body:} If you improve inventory turn but pay more. in freight costs for multiple shipments or your warehouse has to increase their variable costs. to process the additional shipments, the net result may be a loss. 4. An upsell feature on the web is a visual reminder of how much money a customer can. spend before the next shipping \& handling threshold is met. King Arthur Flour is an. excellent example of how to improve upsell and increase items per order. Showing the. amount available, relevant. choices within the price. \\ 
& \textbf{Indep-lex's 1st.} \textit{ID:7220016//} \textit{Title:} Upselling \textit{Body:} What is Upselling? Upselling is a sales technique aimed at persuading customers to purchase a more expensive, upgraded or premium version of the chosen item or other add-ons for the purpose of making a larger sale. eCommerce businesses often combine upselling and cross-selling techniques in an attempt to increase order value and maximize profit.  \\ 
& \textbf{Unifie\textsc{r}$_\text{den}$'s 1st.} \textit{ID:8487388//} \textit{Title:} Acronyms \&Abbreviations \textit{Body:} Ups is an open source source-level debugger developed in the late 1980s for Unix and Unix-like systems, originally developed at the University of Kent by Mark Russell. It supports C and C++, and Fortran on some platforms. The last beta release was in 2003. \\ 
& \textbf{Unifie\textsc{r}$_\text{lex}$'s 1st.} \textit{ID:4754301//} \textit{Title:} Upselling: 75 Strategies, Ideas and Examples \textit{Body:} . Upsell Drip Campaign to upsell B2B/Saas solutions. What is it? The upsell for B2B/Saas solutions email is meant to add to the services. These emails offer premium services or upgrades for users on paying, free or trial accounts. When is it sent? Upsell emails for B2B/Saas solutions are meant to extend the usability and functionality of the software.  \\ 
& \textbf{Unifie\textsc{r}$_\text{uni}$'s 1st.} \textit{ID:4754301//} \textit{Title:} Upselling: 75 Strategies, Ideas and Examples \textit{Body:} . Upsell Drip Campaign to upsell B2B/Saas solutions. What is it? The upsell for B2B/Saas solutions email is meant to add to the services. These emails offer premium services or upgrades for users on paying, free or trial accounts. When is it sent? Upsell emails for B2B/Saas solutions are meant to extend the usability and functionality of the software.  \\ 
\bottomrule
\end{tabular}
\end{table*}

%% file: arxiv-unifier.bbl

\begin{thebibliography}{57}

%%% the Web address.

\ifx \showCODEN    \undefined \def \showCODEN     #1{\unskip}     \fi
\ifx \showDOI      \undefined \def \showDOI       #1{#1}\fi
\ifx \showISBNx    \undefined \def \showISBNx     #1{\unskip}     \fi
\ifx \showISBNxiii \undefined \def \showISBNxiii  #1{\unskip}     \fi
\ifx \showISSN     \undefined \def \showISSN      #1{\unskip}     \fi
\ifx \showLCCN     \undefined \def \showLCCN      #1{\unskip}     \fi
\ifx \shownote     \undefined \def \shownote      #1{#1}          \fi
\ifx \showarticletitle \undefined \def \showarticletitle #1{#1}   \fi
\ifx \showURL      \undefined \def \showURL       {\relax}        \fi
% invisible to TeX
\providecommand\bibfield[2]{#2}
\providecommand\bibinfo[2]{#2}
\providecommand\natexlab[1]{#1}
\providecommand\showeprint[2][]{arXiv:#2}

\bibitem[Beltagy et~al\mbox{.}(2020)]%
        {Beltagy2020Longformer}
\bibfield{author}{\bibinfo{person}{Iz Beltagy}, \bibinfo{person}{Matthew~E.
  Peters}, {and} \bibinfo{person}{Arman Cohan}.}
  \bibinfo{year}{2020}\natexlab{}.
\newblock \showarticletitle{Longformer: The Long-Document Transformer}.
\newblock \bibinfo{journal}{\emph{CoRR}}  \bibinfo{volume}{abs/2004.05150}
  (\bibinfo{year}{2020}).
\newblock
\showeprint[arXiv]{2004.05150}
\urldef\tempurl%
\url{https://arxiv.org/abs/2004.05150}
\showURL{%
\tempurl}


\bibitem[Cai et~al\mbox{.}(2021)]%
        {Cai2021IRSurvey}
\bibfield{author}{\bibinfo{person}{Yinqiong Cai}, \bibinfo{person}{Yixing Fan},
  \bibinfo{person}{Jiafeng Guo}, \bibinfo{person}{Fei Sun},
  \bibinfo{person}{Ruqing Zhang}, {and} \bibinfo{person}{Xueqi Cheng}.}
  \bibinfo{year}{2021}\natexlab{}.
\newblock \showarticletitle{Semantic Models for the First-stage Retrieval: {A}
  Comprehensive Review}.
\newblock \bibinfo{journal}{\emph{CoRR}}  \bibinfo{volume}{abs/2103.04831}
  (\bibinfo{year}{2021}).
\newblock
\showeprint[arXiv]{2103.04831}
\urldef\tempurl%
\url{https://arxiv.org/abs/2103.04831}
\showURL{%
\tempurl}


\bibitem[Chen et~al\mbox{.}(2017)]%
        {Chen2017DrQA}
\bibfield{author}{\bibinfo{person}{Danqi Chen}, \bibinfo{person}{Adam Fisch},
  \bibinfo{person}{Jason Weston}, {and} \bibinfo{person}{Antoine Bordes}.}
  \bibinfo{year}{2017}\natexlab{}.
\newblock \showarticletitle{Reading Wikipedia to Answer Open-Domain Questions}.
  In \bibinfo{booktitle}{\emph{Proceedings of the 55th Annual Meeting of the
  Association for Computational Linguistics, {ACL} 2017, Vancouver, Canada,
  July 30 - August 4, Volume 1: Long Papers}},
  \bibfield{editor}{\bibinfo{person}{Regina Barzilay} {and}
  \bibinfo{person}{Min{-}Yen Kan}} (Eds.). \bibinfo{publisher}{Association for
  Computational Linguistics}, \bibinfo{pages}{1870--1879}.
\newblock
\urldef\tempurl%
\url{https://doi.org/10.18653/v1/P17-1171}
\showDOI{\tempurl}


\bibitem[Chen et~al\mbox{.}(2021b)]%
        {Chen2020HiddenCut}
\bibfield{author}{\bibinfo{person}{Jiaao Chen}, \bibinfo{person}{Dinghan Shen},
  \bibinfo{person}{Weizhu Chen}, {and} \bibinfo{person}{Diyi Yang}.}
  \bibinfo{year}{2021}\natexlab{b}.
\newblock \showarticletitle{HiddenCut: Simple Data Augmentation for Natural
  Language Understanding with Better Generalizability}. In
  \bibinfo{booktitle}{\emph{Proceedings of the 59th Annual Meeting of the
  Association for Computational Linguistics and the 11th International Joint
  Conference on Natural Language Processing, {ACL/IJCNLP} 2021, (Volume 1: Long
  Papers), Virtual Event, August 1-6, 2021}},
  \bibfield{editor}{\bibinfo{person}{Chengqing Zong}, \bibinfo{person}{Fei
  Xia}, \bibinfo{person}{Wenjie Li}, {and} \bibinfo{person}{Roberto Navigli}}
  (Eds.). \bibinfo{publisher}{Association for Computational Linguistics},
  \bibinfo{pages}{4380--4390}.
\newblock
\urldef\tempurl%
\url{https://doi.org/10.18653/v1/2021.acl-long.338}
\showDOI{\tempurl}


\bibitem[Chen et~al\mbox{.}(2021a)]%
        {Chen2021ImitateSparse}
\bibfield{author}{\bibinfo{person}{Xilun Chen}, \bibinfo{person}{Kushal
  Lakhotia}, \bibinfo{person}{Barlas Oguz}, \bibinfo{person}{Anchit Gupta},
  \bibinfo{person}{Patrick S.~H. Lewis}, \bibinfo{person}{Stan Peshterliev},
  \bibinfo{person}{Yashar Mehdad}, \bibinfo{person}{Sonal Gupta}, {and}
  \bibinfo{person}{Wen{-}tau Yih}.} \bibinfo{year}{2021}\natexlab{a}.
\newblock \showarticletitle{Salient Phrase Aware Dense Retrieval: Can a Dense
  Retriever Imitate a Sparse One?}
\newblock \bibinfo{journal}{\emph{CoRR}}  \bibinfo{volume}{abs/2110.06918}
  (\bibinfo{year}{2021}).
\newblock
\showeprint[arXiv]{2110.06918}
\urldef\tempurl%
\url{https://arxiv.org/abs/2110.06918}
\showURL{%
\tempurl}


\bibitem[Chuang et~al\mbox{.}(2022)]%
        {Chuang2022DiffCSE}
\bibfield{author}{\bibinfo{person}{Yung{-}Sung Chuang}, \bibinfo{person}{Rumen
  Dangovski}, \bibinfo{person}{Hongyin Luo}, \bibinfo{person}{Yang Zhang},
  \bibinfo{person}{Shiyu Chang}, \bibinfo{person}{Marin Soljacic},
  \bibinfo{person}{Shang{-}Wen Li}, \bibinfo{person}{Wen{-}tau Yih},
  \bibinfo{person}{Yoon Kim}, {and} \bibinfo{person}{James~R. Glass}.}
  \bibinfo{year}{2022}\natexlab{}.
\newblock \showarticletitle{DiffCSE: Difference-based Contrastive Learning for
  Sentence Embeddings}.
\newblock \bibinfo{journal}{\emph{CoRR}}  \bibinfo{volume}{abs/2204.10298}
  (\bibinfo{year}{2022}).
\newblock
\urldef\tempurl%
\url{https://doi.org/10.48550/arXiv.2204.10298}
\showDOI{\tempurl}
\showeprint[arXiv]{2204.10298}


\bibitem[Craswell et~al\mbox{.}(2020)]%
        {Craswell2020TREC19}
\bibfield{author}{\bibinfo{person}{Nick Craswell}, \bibinfo{person}{Bhaskar
  Mitra}, \bibinfo{person}{Emine Yilmaz}, \bibinfo{person}{Daniel Campos},
  {and} \bibinfo{person}{Ellen~M. Voorhees}.} \bibinfo{year}{2020}\natexlab{}.
\newblock \showarticletitle{Overview of the {TREC} 2019 deep learning track}.
\newblock \bibinfo{journal}{\emph{CoRR}}  \bibinfo{volume}{abs/2003.07820}
  (\bibinfo{year}{2020}).
\newblock
\showeprint[arXiv]{2003.07820}
\urldef\tempurl%
\url{https://arxiv.org/abs/2003.07820}
\showURL{%
\tempurl}


\bibitem[Dai and Callan(2019)]%
        {Dai2019DeepCT}
\bibfield{author}{\bibinfo{person}{Zhuyun Dai} {and} \bibinfo{person}{Jamie
  Callan}.} \bibinfo{year}{2019}\natexlab{}.
\newblock \showarticletitle{Context-Aware Sentence/Passage Term Importance
  Estimation For First Stage Retrieval}.
\newblock \bibinfo{journal}{\emph{CoRR}}  \bibinfo{volume}{abs/1910.10687}
  (\bibinfo{year}{2019}).
\newblock
\showeprint[arXiv]{1910.10687}
\urldef\tempurl%
\url{http://arxiv.org/abs/1910.10687}
\showURL{%
\tempurl}


\bibitem[Devlin et~al\mbox{.}(2019)]%
        {Devlin2019BERT}
\bibfield{author}{\bibinfo{person}{Jacob Devlin}, \bibinfo{person}{Ming{-}Wei
  Chang}, \bibinfo{person}{Kenton Lee}, {and} \bibinfo{person}{Kristina
  Toutanova}.} \bibinfo{year}{2019}\natexlab{}.
\newblock \showarticletitle{{BERT:} Pre-training of Deep Bidirectional
  Transformers for Language Understanding}. In
  \bibinfo{booktitle}{\emph{Proceedings of the 2019 Conference of the North
  American Chapter of the Association for Computational Linguistics: Human
  Language Technologies, {NAACL-HLT} 2019, Minneapolis, MN, USA, June 2-7,
  2019, Volume 1 (Long and Short Papers)}},
  \bibfield{editor}{\bibinfo{person}{Jill Burstein}, \bibinfo{person}{Christy
  Doran}, {and} \bibinfo{person}{Thamar Solorio}} (Eds.).
  \bibinfo{publisher}{Association for Computational Linguistics},
  \bibinfo{pages}{4171--4186}.
\newblock
\urldef\tempurl%
\url{https://doi.org/10.18653/v1/n19-1423}
\showDOI{\tempurl}


\bibitem[Formal et~al\mbox{.}(2021a)]%
        {Formal2021SPLADEv2}
\bibfield{author}{\bibinfo{person}{Thibault Formal}, \bibinfo{person}{Carlos
  Lassance}, \bibinfo{person}{Benjamin Piwowarski}, {and}
  \bibinfo{person}{St{\'{e}}phane Clinchant}.}
  \bibinfo{year}{2021}\natexlab{a}.
\newblock \showarticletitle{{SPLADE} v2: Sparse Lexical and Expansion Model for
  Information Retrieval}.
\newblock \bibinfo{journal}{\emph{CoRR}}  \bibinfo{volume}{abs/2109.10086}
  (\bibinfo{year}{2021}).
\newblock
\showeprint[arXiv]{2109.10086}
\urldef\tempurl%
\url{https://arxiv.org/abs/2109.10086}
\showURL{%
\tempurl}


\bibitem[Formal et~al\mbox{.}(2022)]%
        {Formal2022SPLADE++}
\bibfield{author}{\bibinfo{person}{Thibault Formal}, \bibinfo{person}{Carlos
  Lassance}, \bibinfo{person}{Benjamin Piwowarski}, {and}
  \bibinfo{person}{St{\'{e}}phane Clinchant}.} \bibinfo{year}{2022}\natexlab{}.
\newblock \showarticletitle{From Distillation to Hard Negative Sampling: Making
  Sparse Neural {IR} Models More Effective}.
\newblock \bibinfo{journal}{\emph{CoRR}}  \bibinfo{volume}{abs/2205.04733}
  (\bibinfo{year}{2022}).
\newblock
\urldef\tempurl%
\url{https://doi.org/10.48550/arXiv.2205.04733}
\showDOI{\tempurl}
\showeprint[arXiv]{2205.04733}


\bibitem[Formal et~al\mbox{.}(2021b)]%
        {Formal2021SPLADE}
\bibfield{author}{\bibinfo{person}{Thibault Formal}, \bibinfo{person}{Benjamin
  Piwowarski}, {and} \bibinfo{person}{St{\'{e}}phane Clinchant}.}
  \bibinfo{year}{2021}\natexlab{b}.
\newblock \showarticletitle{{SPLADE:} Sparse Lexical and Expansion Model for
  First Stage Ranking}. In \bibinfo{booktitle}{\emph{{SIGIR} '21: The 44th
  International {ACM} {SIGIR} Conference on Research and Development in
  Information Retrieval, Virtual Event, Canada, July 11-15, 2021}},
  \bibfield{editor}{\bibinfo{person}{Fernando Diaz}, \bibinfo{person}{Chirag
  Shah}, \bibinfo{person}{Torsten Suel}, \bibinfo{person}{Pablo Castells},
  \bibinfo{person}{Rosie Jones}, {and} \bibinfo{person}{Tetsuya Sakai}} (Eds.).
  \bibinfo{publisher}{{ACM}}, \bibinfo{pages}{2288--2292}.
\newblock
\urldef\tempurl%
\url{https://doi.org/10.1145/3404835.3463098}
\showDOI{\tempurl}


\bibitem[Gao and Callan(2021a)]%
        {Gao2021Condenser}
\bibfield{author}{\bibinfo{person}{Luyu Gao} {and} \bibinfo{person}{Jamie
  Callan}.} \bibinfo{year}{2021}\natexlab{a}.
\newblock \showarticletitle{Condenser: a Pre-training Architecture for Dense
  Retrieval}. In \bibinfo{booktitle}{\emph{Proceedings of the 2021 Conference
  on Empirical Methods in Natural Language Processing, {EMNLP} 2021, Virtual
  Event / Punta Cana, Dominican Republic, 7-11 November, 2021}},
  \bibfield{editor}{\bibinfo{person}{Marie{-}Francine Moens},
  \bibinfo{person}{Xuanjing Huang}, \bibinfo{person}{Lucia Specia}, {and}
  \bibinfo{person}{Scott~Wen{-}tau Yih}} (Eds.).
  \bibinfo{publisher}{Association for Computational Linguistics},
  \bibinfo{pages}{981--993}.
\newblock
\urldef\tempurl%
\url{https://doi.org/10.18653/v1/2021.emnlp-main.75}
\showDOI{\tempurl}


\bibitem[Gao and Callan(2021b)]%
        {Gao2021coCondenser}
\bibfield{author}{\bibinfo{person}{Luyu Gao} {and} \bibinfo{person}{Jamie
  Callan}.} \bibinfo{year}{2021}\natexlab{b}.
\newblock \showarticletitle{Unsupervised Corpus Aware Language Model
  Pre-training for Dense Passage Retrieval}.
\newblock \bibinfo{journal}{\emph{CoRR}}  \bibinfo{volume}{abs/2108.05540}
  (\bibinfo{year}{2021}).
\newblock
\showeprint[arXiv]{2108.05540}
\urldef\tempurl%
\url{https://arxiv.org/abs/2108.05540}
\showURL{%
\tempurl}


\bibitem[Gao et~al\mbox{.}(2021a)]%
        {Gao2021COIL}
\bibfield{author}{\bibinfo{person}{Luyu Gao}, \bibinfo{person}{Zhuyun Dai},
  {and} \bibinfo{person}{Jamie Callan}.} \bibinfo{year}{2021}\natexlab{a}.
\newblock \showarticletitle{{COIL:} Revisit Exact Lexical Match in Information
  Retrieval with Contextualized Inverted List}. In
  \bibinfo{booktitle}{\emph{Proceedings of the 2021 Conference of the North
  American Chapter of the Association for Computational Linguistics: Human
  Language Technologies, {NAACL-HLT} 2021, Online, June 6-11, 2021}},
  \bibfield{editor}{\bibinfo{person}{Kristina Toutanova}, \bibinfo{person}{Anna
  Rumshisky}, \bibinfo{person}{Luke Zettlemoyer}, \bibinfo{person}{Dilek
  Hakkani{-}T{\"{u}}r}, \bibinfo{person}{Iz~Beltagy}, \bibinfo{person}{Steven
  Bethard}, \bibinfo{person}{Ryan Cotterell}, \bibinfo{person}{Tanmoy
  Chakraborty}, {and} \bibinfo{person}{Yichao Zhou}} (Eds.).
  \bibinfo{publisher}{Association for Computational Linguistics},
  \bibinfo{pages}{3030--3042}.
\newblock
\urldef\tempurl%
\url{https://doi.org/10.18653/v1/2021.naacl-main.241}
\showDOI{\tempurl}


\bibitem[Gao et~al\mbox{.}(2021b)]%
        {Gao2021CLEAR}
\bibfield{author}{\bibinfo{person}{Luyu Gao}, \bibinfo{person}{Zhuyun Dai},
  \bibinfo{person}{Tongfei Chen}, \bibinfo{person}{Zhen Fan},
  \bibinfo{person}{Benjamin~Van Durme}, {and} \bibinfo{person}{Jamie Callan}.}
  \bibinfo{year}{2021}\natexlab{b}.
\newblock \showarticletitle{Complement Lexical Retrieval Model with Semantic
  Residual Embeddings}. In \bibinfo{booktitle}{\emph{Advances in Information
  Retrieval - 43rd European Conference on {IR} Research, {ECIR} 2021, Virtual
  Event, March 28 - April 1, 2021, Proceedings, Part {I}}}
  \emph{(\bibinfo{series}{Lecture Notes in Computer Science},
  Vol.~\bibinfo{volume}{12656})}, \bibfield{editor}{\bibinfo{person}{Djoerd
  Hiemstra}, \bibinfo{person}{Marie{-}Francine Moens}, \bibinfo{person}{Josiane
  Mothe}, \bibinfo{person}{Raffaele Perego}, \bibinfo{person}{Martin Potthast},
  {and} \bibinfo{person}{Fabrizio Sebastiani}} (Eds.).
  \bibinfo{publisher}{Springer}, \bibinfo{pages}{146--160}.
\newblock
\urldef\tempurl%
\url{https://doi.org/10.1007/978-3-030-72113-8\_10}
\showDOI{\tempurl}


\bibitem[Gao et~al\mbox{.}(2021c)]%
        {Gao2021SimCSE}
\bibfield{author}{\bibinfo{person}{Tianyu Gao}, \bibinfo{person}{Xingcheng
  Yao}, {and} \bibinfo{person}{Danqi Chen}.} \bibinfo{year}{2021}\natexlab{c}.
\newblock \showarticletitle{SimCSE: Simple Contrastive Learning of Sentence
  Embeddings}. In \bibinfo{booktitle}{\emph{Proceedings of the 2021 Conference
  on Empirical Methods in Natural Language Processing, {EMNLP} 2021, Virtual
  Event / Punta Cana, Dominican Republic, 7-11 November, 2021}},
  \bibfield{editor}{\bibinfo{person}{Marie{-}Francine Moens},
  \bibinfo{person}{Xuanjing Huang}, \bibinfo{person}{Lucia Specia}, {and}
  \bibinfo{person}{Scott~Wen{-}tau Yih}} (Eds.).
  \bibinfo{publisher}{Association for Computational Linguistics},
  \bibinfo{pages}{6894--6910}.
\newblock
\urldef\tempurl%
\url{https://doi.org/10.18653/v1/2021.emnlp-main.552}
\showDOI{\tempurl}


\bibitem[Han et~al\mbox{.}(2018)]%
        {Han2018coteaching}
\bibfield{author}{\bibinfo{person}{Bo Han}, \bibinfo{person}{Quanming Yao},
  \bibinfo{person}{Xingrui Yu}, \bibinfo{person}{Gang Niu},
  \bibinfo{person}{Miao Xu}, \bibinfo{person}{Weihua Hu},
  \bibinfo{person}{Ivor~W. Tsang}, {and} \bibinfo{person}{Masashi Sugiyama}.}
  \bibinfo{year}{2018}\natexlab{}.
\newblock \showarticletitle{Co-teaching: Robust training of deep neural
  networks with extremely noisy labels}. In \bibinfo{booktitle}{\emph{Advances
  in Neural Information Processing Systems 31: Annual Conference on Neural
  Information Processing Systems 2018, NeurIPS 2018, December 3-8, 2018,
  Montr{\'{e}}al, Canada}}, \bibfield{editor}{\bibinfo{person}{Samy Bengio},
  \bibinfo{person}{Hanna~M. Wallach}, \bibinfo{person}{Hugo Larochelle},
  \bibinfo{person}{Kristen Grauman}, \bibinfo{person}{Nicol{\`{o}}
  Cesa{-}Bianchi}, {and} \bibinfo{person}{Roman Garnett}} (Eds.).
  \bibinfo{pages}{8536--8546}.
\newblock
\urldef\tempurl%
\url{https://proceedings.neurips.cc/paper/2018/hash/a19744e268754fb0148b017647355b7b-Abstract.html}
\showURL{%
\tempurl}


\bibitem[Hofst{\"{a}}tter et~al\mbox{.}(2020)]%
        {Hofstatter2020Margin-MSE}
\bibfield{author}{\bibinfo{person}{Sebastian Hofst{\"{a}}tter},
  \bibinfo{person}{Sophia Althammer}, \bibinfo{person}{Michael Schr{\"{o}}der},
  \bibinfo{person}{Mete Sertkan}, {and} \bibinfo{person}{Allan Hanbury}.}
  \bibinfo{year}{2020}\natexlab{}.
\newblock \showarticletitle{Improving Efficient Neural Ranking Models with
  Cross-Architecture Knowledge Distillation}.
\newblock \bibinfo{journal}{\emph{CoRR}}  \bibinfo{volume}{abs/2010.02666}
  (\bibinfo{year}{2020}).
\newblock
\showeprint[arXiv]{2010.02666}
\urldef\tempurl%
\url{https://arxiv.org/abs/2010.02666}
\showURL{%
\tempurl}


\bibitem[Hofst{\"{a}}tter et~al\mbox{.}(2021)]%
        {Hofstatter2021TAS-B}
\bibfield{author}{\bibinfo{person}{Sebastian Hofst{\"{a}}tter},
  \bibinfo{person}{Sheng{-}Chieh Lin}, \bibinfo{person}{Jheng{-}Hong Yang},
  \bibinfo{person}{Jimmy Lin}, {and} \bibinfo{person}{Allan Hanbury}.}
  \bibinfo{year}{2021}\natexlab{}.
\newblock \showarticletitle{Efficiently Teaching an Effective Dense Retriever
  with Balanced Topic Aware Sampling}. In \bibinfo{booktitle}{\emph{{SIGIR}
  '21: The 44th International {ACM} {SIGIR} Conference on Research and
  Development in Information Retrieval, Virtual Event, Canada, July 11-15,
  2021}}, \bibfield{editor}{\bibinfo{person}{Fernando Diaz},
  \bibinfo{person}{Chirag Shah}, \bibinfo{person}{Torsten Suel},
  \bibinfo{person}{Pablo Castells}, \bibinfo{person}{Rosie Jones}, {and}
  \bibinfo{person}{Tetsuya Sakai}} (Eds.). \bibinfo{publisher}{{ACM}},
  \bibinfo{pages}{113--122}.
\newblock
\urldef\tempurl%
\url{https://doi.org/10.1145/3404835.3462891}
\showDOI{\tempurl}


\bibitem[Izacard et~al\mbox{.}(2021)]%
        {Izacard2021Contriever}
\bibfield{author}{\bibinfo{person}{Gautier Izacard}, \bibinfo{person}{Mathilde
  Caron}, \bibinfo{person}{Lucas Hosseini}, \bibinfo{person}{Sebastian Riedel},
  \bibinfo{person}{Piotr Bojanowski}, \bibinfo{person}{Armand Joulin}, {and}
  \bibinfo{person}{Edouard Grave}.} \bibinfo{year}{2021}\natexlab{}.
\newblock \showarticletitle{Towards Unsupervised Dense Information Retrieval
  with Contrastive Learning}.
\newblock \bibinfo{journal}{\emph{CoRR}}  \bibinfo{volume}{abs/2112.09118}
  (\bibinfo{year}{2021}).
\newblock
\showeprint[arXiv]{2112.09118}
\urldef\tempurl%
\url{https://arxiv.org/abs/2112.09118}
\showURL{%
\tempurl}


\bibitem[Jin et~al\mbox{.}(2022)]%
        {Jin2022InstanceAwarePrompt}
\bibfield{author}{\bibinfo{person}{Feihu Jin}, \bibinfo{person}{Jinliang Lu},
  \bibinfo{person}{Jiajun Zhang}, {and} \bibinfo{person}{Chengqing Zong}.}
  \bibinfo{year}{2022}\natexlab{}.
\newblock \showarticletitle{Instance-aware Prompt Learning for Language
  Understanding and Generation}.
\newblock \bibinfo{journal}{\emph{CoRR}}  \bibinfo{volume}{abs/2201.07126}
  (\bibinfo{year}{2022}).
\newblock
\showeprint[arXiv]{2201.07126}
\urldef\tempurl%
\url{https://arxiv.org/abs/2201.07126}
\showURL{%
\tempurl}


\bibitem[Karpukhin et~al\mbox{.}(2020)]%
        {Karpukhin2020DPR}
\bibfield{author}{\bibinfo{person}{Vladimir Karpukhin}, \bibinfo{person}{Barlas
  Oguz}, \bibinfo{person}{Sewon Min}, \bibinfo{person}{Patrick S.~H. Lewis},
  \bibinfo{person}{Ledell Wu}, \bibinfo{person}{Sergey Edunov},
  \bibinfo{person}{Danqi Chen}, {and} \bibinfo{person}{Wen{-}tau Yih}.}
  \bibinfo{year}{2020}\natexlab{}.
\newblock \showarticletitle{Dense Passage Retrieval for Open-Domain Question
  Answering}. In \bibinfo{booktitle}{\emph{Proceedings of the 2020 Conference
  on Empirical Methods in Natural Language Processing, {EMNLP} 2020, Online,
  November 16-20, 2020}}, \bibfield{editor}{\bibinfo{person}{Bonnie Webber},
  \bibinfo{person}{Trevor Cohn}, \bibinfo{person}{Yulan He}, {and}
  \bibinfo{person}{Yang Liu}} (Eds.). \bibinfo{publisher}{Association for
  Computational Linguistics}, \bibinfo{pages}{6769--6781}.
\newblock
\urldef\tempurl%
\url{https://doi.org/10.18653/v1/2020.emnlp-main.550}
\showDOI{\tempurl}


\bibitem[Khattab and Zaharia(2020)]%
        {Khattab2020ColBERT}
\bibfield{author}{\bibinfo{person}{Omar Khattab} {and} \bibinfo{person}{Matei
  Zaharia}.} \bibinfo{year}{2020}\natexlab{}.
\newblock \showarticletitle{ColBERT: Efficient and Effective Passage Search via
  Contextualized Late Interaction over {BERT}}. In
  \bibinfo{booktitle}{\emph{Proceedings of the 43rd International {ACM} {SIGIR}
  conference on research and development in Information Retrieval, {SIGIR}
  2020, Virtual Event, China, July 25-30, 2020}},
  \bibfield{editor}{\bibinfo{person}{Jimmy Huang}, \bibinfo{person}{Yi~Chang},
  \bibinfo{person}{Xueqi Cheng}, \bibinfo{person}{Jaap Kamps},
  \bibinfo{person}{Vanessa Murdock}, \bibinfo{person}{Ji{-}Rong Wen}, {and}
  \bibinfo{person}{Yiqun Liu}} (Eds.). \bibinfo{publisher}{{ACM}},
  \bibinfo{pages}{39--48}.
\newblock
\urldef\tempurl%
\url{https://doi.org/10.1145/3397271.3401075}
\showDOI{\tempurl}


\bibitem[Kuzi et~al\mbox{.}(2020)]%
        {Kuzi2020ScoreAgg}
\bibfield{author}{\bibinfo{person}{Saar Kuzi}, \bibinfo{person}{Mingyang
  Zhang}, \bibinfo{person}{Cheng Li}, \bibinfo{person}{Michael Bendersky},
  {and} \bibinfo{person}{Marc Najork}.} \bibinfo{year}{2020}\natexlab{}.
\newblock \showarticletitle{Leveraging Semantic and Lexical Matching to Improve
  the Recall of Document Retrieval Systems: {A} Hybrid Approach}.
\newblock \bibinfo{journal}{\emph{CoRR}}  \bibinfo{volume}{abs/2010.01195}
  (\bibinfo{year}{2020}).
\newblock
\showeprint[arXiv]{2010.01195}
\urldef\tempurl%
\url{https://arxiv.org/abs/2010.01195}
\showURL{%
\tempurl}


\bibitem[Lee et~al\mbox{.}(2020)]%
        {Lee2020SPARC}
\bibfield{author}{\bibinfo{person}{Jinhyuk Lee}, \bibinfo{person}{Min~Joon
  Seo}, \bibinfo{person}{Hannaneh Hajishirzi}, {and} \bibinfo{person}{Jaewoo
  Kang}.} \bibinfo{year}{2020}\natexlab{}.
\newblock \showarticletitle{Contextualized Sparse Representations for Real-Time
  Open-Domain Question Answering}. In \bibinfo{booktitle}{\emph{Proceedings of
  the 58th Annual Meeting of the Association for Computational Linguistics,
  {ACL} 2020, Online, July 5-10, 2020}}, \bibfield{editor}{\bibinfo{person}{Dan
  Jurafsky}, \bibinfo{person}{Joyce Chai}, \bibinfo{person}{Natalie Schluter},
  {and} \bibinfo{person}{Joel~R. Tetreault}} (Eds.).
  \bibinfo{publisher}{Association for Computational Linguistics},
  \bibinfo{pages}{912--919}.
\newblock
\urldef\tempurl%
\url{https://doi.org/10.18653/v1/2020.acl-main.85}
\showDOI{\tempurl}


\bibitem[Liang et~al\mbox{.}(2021)]%
        {Liang2021rdrop}
\bibfield{author}{\bibinfo{person}{Xiaobo Liang}, \bibinfo{person}{Lijun Wu},
  \bibinfo{person}{Juntao Li}, \bibinfo{person}{Yue Wang}, \bibinfo{person}{Qi
  Meng}, \bibinfo{person}{Tao Qin}, \bibinfo{person}{Wei Chen},
  \bibinfo{person}{Min Zhang}, {and} \bibinfo{person}{Tie{-}Yan Liu}.}
  \bibinfo{year}{2021}\natexlab{}.
\newblock \showarticletitle{R-Drop: Regularized Dropout for Neural Networks}.
  In \bibinfo{booktitle}{\emph{Advances in Neural Information Processing
  Systems 34: Annual Conference on Neural Information Processing Systems 2021,
  NeurIPS 2021, December 6-14, 2021, virtual}},
  \bibfield{editor}{\bibinfo{person}{Marc'Aurelio Ranzato},
  \bibinfo{person}{Alina Beygelzimer}, \bibinfo{person}{Yann~N. Dauphin},
  \bibinfo{person}{Percy Liang}, {and} \bibinfo{person}{Jennifer~Wortman
  Vaughan}} (Eds.). \bibinfo{pages}{10890--10905}.
\newblock
\urldef\tempurl%
\url{https://proceedings.neurips.cc/paper/2021/hash/5a66b9200f29ac3fa0ae244cc2a51b39-Abstract.html}
\showURL{%
\tempurl}


\bibitem[Lin and Ma(2021)]%
        {Lin2021UniCOIL}
\bibfield{author}{\bibinfo{person}{Jimmy Lin} {and} \bibinfo{person}{Xueguang
  Ma}.} \bibinfo{year}{2021}\natexlab{}.
\newblock \showarticletitle{A Few Brief Notes on DeepImpact, COIL, and a
  Conceptual Framework for Information Retrieval Techniques}.
\newblock \bibinfo{journal}{\emph{CoRR}}  \bibinfo{volume}{abs/2106.14807}
  (\bibinfo{year}{2021}).
\newblock
\showeprint[arXiv]{2106.14807}
\urldef\tempurl%
\url{https://arxiv.org/abs/2106.14807}
\showURL{%
\tempurl}


\bibitem[Lin et~al\mbox{.}(2021)]%
        {Lin2021TCT}
\bibfield{author}{\bibinfo{person}{Sheng-Chieh Lin},
  \bibinfo{person}{Jheng-Hong Yang}, {and} \bibinfo{person}{Jimmy Lin}.}
  \bibinfo{year}{2021}\natexlab{}.
\newblock \showarticletitle{In-Batch Negatives for Knowledge Distillation with
  Tightly-Coupled Teachers for Dense Retrieval}. In
  \bibinfo{booktitle}{\emph{Proceedings of the 6th Workshop on Representation
  Learning for NLP (RepL4NLP-2021)}}. \bibinfo{publisher}{Association for
  Computational Linguistics}, \bibinfo{address}{Online},
  \bibinfo{pages}{163--173}.
\newblock
\urldef\tempurl%
\url{https://doi.org/10.18653/v1/2021.repl4nlp-1.17}
\showDOI{\tempurl}


\bibitem[Liu et~al\mbox{.}(2019)]%
        {Liu2019RoBERTa}
\bibfield{author}{\bibinfo{person}{Yinhan Liu}, \bibinfo{person}{Myle Ott},
  \bibinfo{person}{Naman Goyal}, \bibinfo{person}{Jingfei Du},
  \bibinfo{person}{Mandar Joshi}, \bibinfo{person}{Danqi Chen},
  \bibinfo{person}{Omer Levy}, \bibinfo{person}{Mike Lewis},
  \bibinfo{person}{Luke Zettlemoyer}, {and} \bibinfo{person}{Veselin
  Stoyanov}.} \bibinfo{year}{2019}\natexlab{}.
\newblock \showarticletitle{RoBERTa: {A} Robustly Optimized {BERT} Pretraining
  Approach}.
\newblock \bibinfo{journal}{\emph{CoRR}}  \bibinfo{volume}{abs/1907.11692}
  (\bibinfo{year}{2019}).
\newblock
\showeprint[arXiv]{1907.11692}
\urldef\tempurl%
\url{http://arxiv.org/abs/1907.11692}
\showURL{%
\tempurl}


\bibitem[Lu et~al\mbox{.}(2021)]%
        {Lu2021SeedEncoder}
\bibfield{author}{\bibinfo{person}{Shuqi Lu}, \bibinfo{person}{Chenyan Xiong},
  \bibinfo{person}{Di He}, \bibinfo{person}{Guolin Ke}, \bibinfo{person}{Waleed
  Malik}, \bibinfo{person}{Zhicheng Dou}, \bibinfo{person}{Paul Bennett},
  \bibinfo{person}{Tie{-}Yan Liu}, {and} \bibinfo{person}{Arnold Overwijk}.}
  \bibinfo{year}{2021}\natexlab{}.
\newblock \showarticletitle{Less is More: Pre-training a Strong Siamese Encoder
  Using a Weak Decoder}.
\newblock \bibinfo{journal}{\emph{CoRR}}  \bibinfo{volume}{abs/2102.09206}
  (\bibinfo{year}{2021}).
\newblock
\showeprint[arXiv]{2102.09206}
\urldef\tempurl%
\url{https://arxiv.org/abs/2102.09206}
\showURL{%
\tempurl}


\bibitem[Menon et~al\mbox{.}(2021)]%
        {menon2021indefense}
\bibfield{author}{\bibinfo{person}{Aditya~Krishna Menon},
  \bibinfo{person}{Sadeep Jayasumana}, \bibinfo{person}{Seungyeon Kim},
  \bibinfo{person}{Ankit~Singh Rawat}, \bibinfo{person}{Sashank~J Reddi}, {and}
  \bibinfo{person}{Sanjiv Kumar}.} \bibinfo{year}{2021}\natexlab{}.
\newblock \showarticletitle{In defense of dual-encoders for neural ranking}.
\newblock  (\bibinfo{year}{2021}).
\newblock


\bibitem[Muennighoff(2022)]%
        {Muennighoff2022SGPT}
\bibfield{author}{\bibinfo{person}{Niklas Muennighoff}.}
  \bibinfo{year}{2022}\natexlab{}.
\newblock \showarticletitle{{SGPT:} {GPT} Sentence Embeddings for Semantic
  Search}.
\newblock \bibinfo{journal}{\emph{CoRR}}  \bibinfo{volume}{abs/2202.08904}
  (\bibinfo{year}{2022}).
\newblock
\showeprint[arXiv]{2202.08904}
\urldef\tempurl%
\url{https://arxiv.org/abs/2202.08904}
\showURL{%
\tempurl}


\bibitem[Nguyen et~al\mbox{.}(2016)]%
        {Nguyen2016MSMARCO}
\bibfield{author}{\bibinfo{person}{Tri Nguyen}, \bibinfo{person}{Mir
  Rosenberg}, \bibinfo{person}{Xia Song}, \bibinfo{person}{Jianfeng Gao},
  \bibinfo{person}{Saurabh Tiwary}, \bibinfo{person}{Rangan Majumder}, {and}
  \bibinfo{person}{Li Deng}.} \bibinfo{year}{2016}\natexlab{}.
\newblock \showarticletitle{{MS} {MARCO:} {A} Human Generated MAchine Reading
  COmprehension Dataset}. In \bibinfo{booktitle}{\emph{Proceedings of the
  Workshop on Cognitive Computation: Integrating neural and symbolic approaches
  2016 co-located with the 30th Annual Conference on Neural Information
  Processing Systems {(NIPS} 2016), Barcelona, Spain, December 9, 2016}}
  \emph{(\bibinfo{series}{{CEUR} Workshop Proceedings},
  Vol.~\bibinfo{volume}{1773})},
  \bibfield{editor}{\bibinfo{person}{Tarek~Richard Besold},
  \bibinfo{person}{Antoine Bordes}, \bibinfo{person}{Artur~S. d'Avila Garcez},
  {and} \bibinfo{person}{Greg Wayne}} (Eds.). \bibinfo{publisher}{CEUR-WS.org}.
\newblock
\urldef\tempurl%
\url{http://ceur-ws.org/Vol-1773/CoCoNIPS\_2016\_paper9.pdf}
\showURL{%
\tempurl}


\bibitem[Ni et~al\mbox{.}(2021)]%
        {Ni2021GTR}
\bibfield{author}{\bibinfo{person}{Jianmo Ni}, \bibinfo{person}{Chen Qu},
  \bibinfo{person}{Jing Lu}, \bibinfo{person}{Zhuyun Dai},
  \bibinfo{person}{Gustavo~Hern{\'{a}}ndez {\'{A}}brego}, \bibinfo{person}{Ji
  Ma}, \bibinfo{person}{Vincent~Y. Zhao}, \bibinfo{person}{Yi Luan},
  \bibinfo{person}{Keith~B. Hall}, \bibinfo{person}{Ming{-}Wei Chang}, {and}
  \bibinfo{person}{Yinfei Yang}.} \bibinfo{year}{2021}\natexlab{}.
\newblock \showarticletitle{Large Dual Encoders Are Generalizable Retrievers}.
\newblock \bibinfo{journal}{\emph{CoRR}}  \bibinfo{volume}{abs/2112.07899}
  (\bibinfo{year}{2021}).
\newblock
\showeprint[arXiv]{2112.07899}
\urldef\tempurl%
\url{https://arxiv.org/abs/2112.07899}
\showURL{%
\tempurl}


\bibitem[Nogueira et~al\mbox{.}(2019)]%
        {Nogueira2019DT5Q}
\bibfield{author}{\bibinfo{person}{Rodrigo Nogueira}, \bibinfo{person}{Jimmy
  Lin}, {and} \bibinfo{person}{AI Epistemic}.} \bibinfo{year}{2019}\natexlab{}.
\newblock \showarticletitle{From doc2query to docTTTTTquery}.
\newblock \bibinfo{journal}{\emph{Online preprint}}  \bibinfo{volume}{6}
  (\bibinfo{year}{2019}).
\newblock


\bibitem[Paria et~al\mbox{.}(2020)]%
        {Paria2020FLOPS}
\bibfield{author}{\bibinfo{person}{Biswajit Paria},
  \bibinfo{person}{Chih{-}Kuan Yeh}, \bibinfo{person}{Ian~En{-}Hsu Yen},
  \bibinfo{person}{Ning Xu}, \bibinfo{person}{Pradeep Ravikumar}, {and}
  \bibinfo{person}{Barnab{\'{a}}s P{\'{o}}czos}.}
  \bibinfo{year}{2020}\natexlab{}.
\newblock \showarticletitle{Minimizing FLOPs to Learn Efficient Sparse
  Representations}. In \bibinfo{booktitle}{\emph{8th International Conference
  on Learning Representations, {ICLR} 2020, Addis Ababa, Ethiopia, April 26-30,
  2020}}. \bibinfo{publisher}{OpenReview.net}.
\newblock
\urldef\tempurl%
\url{https://openreview.net/forum?id=SygpC6Ntvr}
\showURL{%
\tempurl}


\bibitem[Qu et~al\mbox{.}(2021)]%
        {Qu2021RocketQA}
\bibfield{author}{\bibinfo{person}{Yingqi Qu}, \bibinfo{person}{Yuchen Ding},
  \bibinfo{person}{Jing Liu}, \bibinfo{person}{Kai Liu},
  \bibinfo{person}{Ruiyang Ren}, \bibinfo{person}{Wayne~Xin Zhao},
  \bibinfo{person}{Daxiang Dong}, \bibinfo{person}{Hua Wu}, {and}
  \bibinfo{person}{Haifeng Wang}.} \bibinfo{year}{2021}\natexlab{}.
\newblock \showarticletitle{RocketQA: An Optimized Training Approach to Dense
  Passage Retrieval for Open-Domain Question Answering}. In
  \bibinfo{booktitle}{\emph{Proceedings of the 2021 Conference of the North
  American Chapter of the Association for Computational Linguistics: Human
  Language Technologies, {NAACL-HLT} 2021, Online, June 6-11, 2021}},
  \bibfield{editor}{\bibinfo{person}{Kristina Toutanova}, \bibinfo{person}{Anna
  Rumshisky}, \bibinfo{person}{Luke Zettlemoyer}, \bibinfo{person}{Dilek
  Hakkani{-}T{\"{u}}r}, \bibinfo{person}{Iz~Beltagy}, \bibinfo{person}{Steven
  Bethard}, \bibinfo{person}{Ryan Cotterell}, \bibinfo{person}{Tanmoy
  Chakraborty}, {and} \bibinfo{person}{Yichao Zhou}} (Eds.).
  \bibinfo{publisher}{Association for Computational Linguistics},
  \bibinfo{pages}{5835--5847}.
\newblock
\urldef\tempurl%
\url{https://doi.org/10.18653/v1/2021.naacl-main.466}
\showDOI{\tempurl}


\bibitem[Reimers and Gurevych(2019)]%
        {Reimers2019SentenceBERT}
\bibfield{author}{\bibinfo{person}{Nils Reimers} {and} \bibinfo{person}{Iryna
  Gurevych}.} \bibinfo{year}{2019}\natexlab{}.
\newblock \showarticletitle{Sentence-BERT: Sentence Embeddings using Siamese
  BERT-Networks}. In \bibinfo{booktitle}{\emph{Proceedings of the 2019
  Conference on Empirical Methods in Natural Language Processing and the 9th
  International Joint Conference on Natural Language Processing, {EMNLP-IJCNLP}
  2019, Hong Kong, China, November 3-7, 2019}},
  \bibfield{editor}{\bibinfo{person}{Kentaro Inui}, \bibinfo{person}{Jing
  Jiang}, \bibinfo{person}{Vincent Ng}, {and} \bibinfo{person}{Xiaojun Wan}}
  (Eds.). \bibinfo{publisher}{Association for Computational Linguistics},
  \bibinfo{pages}{3980--3990}.
\newblock
\urldef\tempurl%
\url{https://doi.org/10.18653/v1/D19-1410}
\showDOI{\tempurl}


\bibitem[Ren et~al\mbox{.}(2021a)]%
        {Ren2021PAIR}
\bibfield{author}{\bibinfo{person}{Ruiyang Ren}, \bibinfo{person}{Shangwen Lv},
  \bibinfo{person}{Yingqi Qu}, \bibinfo{person}{Jing Liu},
  \bibinfo{person}{Wayne~Xin Zhao}, \bibinfo{person}{Qiaoqiao She},
  \bibinfo{person}{Hua Wu}, \bibinfo{person}{Haifeng Wang}, {and}
  \bibinfo{person}{Ji{-}Rong Wen}.} \bibinfo{year}{2021}\natexlab{a}.
\newblock \showarticletitle{{PAIR:} Leveraging Passage-Centric Similarity
  Relation for Improving Dense Passage Retrieval}. In
  \bibinfo{booktitle}{\emph{Findings of the Association for Computational
  Linguistics: {ACL/IJCNLP} 2021, Online Event, August 1-6, 2021}}
  \emph{(\bibinfo{series}{Findings of {ACL}},
  Vol.~\bibinfo{volume}{{ACL/IJCNLP} 2021})},
  \bibfield{editor}{\bibinfo{person}{Chengqing Zong}, \bibinfo{person}{Fei
  Xia}, \bibinfo{person}{Wenjie Li}, {and} \bibinfo{person}{Roberto Navigli}}
  (Eds.). \bibinfo{publisher}{Association for Computational Linguistics},
  \bibinfo{pages}{2173--2183}.
\newblock
\urldef\tempurl%
\url{https://doi.org/10.18653/v1/2021.findings-acl.191}
\showDOI{\tempurl}


\bibitem[Ren et~al\mbox{.}(2021b)]%
        {Ren2021RocketQAv2}
\bibfield{author}{\bibinfo{person}{Ruiyang Ren}, \bibinfo{person}{Yingqi Qu},
  \bibinfo{person}{Jing Liu}, \bibinfo{person}{Wayne~Xin Zhao},
  \bibinfo{person}{Qiaoqiao She}, \bibinfo{person}{Hua Wu},
  \bibinfo{person}{Haifeng Wang}, {and} \bibinfo{person}{Ji{-}Rong Wen}.}
  \bibinfo{year}{2021}\natexlab{b}.
\newblock \showarticletitle{RocketQAv2: {A} Joint Training Method for Dense
  Passage Retrieval and Passage Re-ranking}. In
  \bibinfo{booktitle}{\emph{Proceedings of the 2021 Conference on Empirical
  Methods in Natural Language Processing, {EMNLP} 2021, Virtual Event / Punta
  Cana, Dominican Republic, 7-11 November, 2021}},
  \bibfield{editor}{\bibinfo{person}{Marie{-}Francine Moens},
  \bibinfo{person}{Xuanjing Huang}, \bibinfo{person}{Lucia Specia}, {and}
  \bibinfo{person}{Scott~Wen{-}tau Yih}} (Eds.).
  \bibinfo{publisher}{Association for Computational Linguistics},
  \bibinfo{pages}{2825--2835}.
\newblock
\urldef\tempurl%
\url{https://doi.org/10.18653/v1/2021.emnlp-main.224}
\showDOI{\tempurl}


\bibitem[Robertson and Zaragoza(2009)]%
        {Robertson2009BM25}
\bibfield{author}{\bibinfo{person}{Stephen~E. Robertson} {and}
  \bibinfo{person}{Hugo Zaragoza}.} \bibinfo{year}{2009}\natexlab{}.
\newblock \showarticletitle{The Probabilistic Relevance Framework: {BM25} and
  Beyond}.
\newblock \bibinfo{journal}{\emph{Found. Trends Inf. Retr.}}
  \bibinfo{volume}{3}, \bibinfo{number}{4} (\bibinfo{year}{2009}),
  \bibinfo{pages}{333--389}.
\newblock
\urldef\tempurl%
\url{https://doi.org/10.1561/1500000019}
\showDOI{\tempurl}


\bibitem[Santhanam et~al\mbox{.}(2021)]%
        {Khattab2021ColBERTv2}
\bibfield{author}{\bibinfo{person}{Keshav Santhanam}, \bibinfo{person}{Omar
  Khattab}, \bibinfo{person}{Jon Saad{-}Falcon}, \bibinfo{person}{Christopher
  Potts}, {and} \bibinfo{person}{Matei Zaharia}.}
  \bibinfo{year}{2021}\natexlab{}.
\newblock \showarticletitle{ColBERTv2: Effective and Efficient Retrieval via
  Lightweight Late Interaction}.
\newblock \bibinfo{journal}{\emph{CoRR}}  \bibinfo{volume}{abs/2112.01488}
  (\bibinfo{year}{2021}).
\newblock
\showeprint[arXiv]{2112.01488}
\urldef\tempurl%
\url{https://arxiv.org/abs/2112.01488}
\showURL{%
\tempurl}


\bibitem[Shen et~al\mbox{.}(2019)]%
        {Shen2019localglobal}
\bibfield{author}{\bibinfo{person}{Tao Shen}, \bibinfo{person}{Tianyi Zhou},
  \bibinfo{person}{Guodong Long}, \bibinfo{person}{Jing Jiang}, {and}
  \bibinfo{person}{Chengqi Zhang}.} \bibinfo{year}{2019}\natexlab{}.
\newblock \showarticletitle{Tensorized Self-Attention: Efficiently Modeling
  Pairwise and Global Dependencies Together}. In
  \bibinfo{booktitle}{\emph{Proceedings of the 2019 Conference of the North
  American Chapter of the Association for Computational Linguistics: Human
  Language Technologies, {NAACL-HLT} 2019, Minneapolis, MN, USA, June 2-7,
  2019, Volume 1 (Long and Short Papers)}},
  \bibfield{editor}{\bibinfo{person}{Jill Burstein}, \bibinfo{person}{Christy
  Doran}, {and} \bibinfo{person}{Thamar Solorio}} (Eds.).
  \bibinfo{publisher}{Association for Computational Linguistics},
  \bibinfo{pages}{1256--1266}.
\newblock
\urldef\tempurl%
\url{https://doi.org/10.18653/v1/n19-1127}
\showDOI{\tempurl}


\bibitem[Song et~al\mbox{.}(2021)]%
        {Song2021wordpiece}
\bibfield{author}{\bibinfo{person}{Xinying Song}, \bibinfo{person}{Alex
  Salcianu}, \bibinfo{person}{Yang Song}, \bibinfo{person}{Dave Dopson}, {and}
  \bibinfo{person}{Denny Zhou}.} \bibinfo{year}{2021}\natexlab{}.
\newblock \showarticletitle{Fast WordPiece Tokenization}. In
  \bibinfo{booktitle}{\emph{Proceedings of the 2021 Conference on Empirical
  Methods in Natural Language Processing, {EMNLP} 2021, Virtual Event / Punta
  Cana, Dominican Republic, 7-11 November, 2021}},
  \bibfield{editor}{\bibinfo{person}{Marie{-}Francine Moens},
  \bibinfo{person}{Xuanjing Huang}, \bibinfo{person}{Lucia Specia}, {and}
  \bibinfo{person}{Scott~Wen{-}tau Yih}} (Eds.).
  \bibinfo{publisher}{Association for Computational Linguistics},
  \bibinfo{pages}{2089--2103}.
\newblock
\urldef\tempurl%
\url{https://doi.org/10.18653/v1/2021.emnlp-main.160}
\showDOI{\tempurl}


\bibitem[Sun et~al\mbox{.}(2019)]%
        {Sun2019RotatE}
\bibfield{author}{\bibinfo{person}{Zhiqing Sun}, \bibinfo{person}{Zhi{-}Hong
  Deng}, \bibinfo{person}{Jian{-}Yun Nie}, {and} \bibinfo{person}{Jian Tang}.}
  \bibinfo{year}{2019}\natexlab{}.
\newblock \showarticletitle{RotatE: Knowledge Graph Embedding by Relational
  Rotation in Complex Space}. In \bibinfo{booktitle}{\emph{7th International
  Conference on Learning Representations, {ICLR} 2019, New Orleans, LA, USA,
  May 6-9, 2019}}. \bibinfo{publisher}{OpenReview.net}.
\newblock
\urldef\tempurl%
\url{https://openreview.net/forum?id=HkgEQnRqYQ}
\showURL{%
\tempurl}


\bibitem[Sutskever et~al\mbox{.}(2014)]%
        {Sutskever2014seq2seq}
\bibfield{author}{\bibinfo{person}{Ilya Sutskever}, \bibinfo{person}{Oriol
  Vinyals}, {and} \bibinfo{person}{Quoc~V. Le}.}
  \bibinfo{year}{2014}\natexlab{}.
\newblock \showarticletitle{Sequence to Sequence Learning with Neural
  Networks}. In \bibinfo{booktitle}{\emph{Advances in Neural Information
  Processing Systems 27: Annual Conference on Neural Information Processing
  Systems 2014, December 8-13 2014, Montreal, Quebec, Canada}},
  \bibfield{editor}{\bibinfo{person}{Zoubin Ghahramani}, \bibinfo{person}{Max
  Welling}, \bibinfo{person}{Corinna Cortes}, \bibinfo{person}{Neil~D.
  Lawrence}, {and} \bibinfo{person}{Kilian~Q. Weinberger}} (Eds.).
  \bibinfo{pages}{3104--3112}.
\newblock
\urldef\tempurl%
\url{https://proceedings.neurips.cc/paper/2014/hash/a14ac55a4f27472c5d894ec1c3c743d2-Abstract.html}
\showURL{%
\tempurl}


\bibitem[Thakur et~al\mbox{.}(2021)]%
        {Thakur2021BEIR}
\bibfield{author}{\bibinfo{person}{Nandan Thakur}, \bibinfo{person}{Nils
  Reimers}, \bibinfo{person}{Andreas R{\"{u}}ckl{\'{e}}},
  \bibinfo{person}{Abhishek Srivastava}, {and} \bibinfo{person}{Iryna
  Gurevych}.} \bibinfo{year}{2021}\natexlab{}.
\newblock \showarticletitle{{BEIR:} {A} Heterogenous Benchmark for Zero-shot
  Evaluation of Information Retrieval Models}.
\newblock \bibinfo{journal}{\emph{CoRR}}  \bibinfo{volume}{abs/2104.08663}
  (\bibinfo{year}{2021}).
\newblock
\showeprint[arXiv]{2104.08663}
\urldef\tempurl%
\url{https://arxiv.org/abs/2104.08663}
\showURL{%
\tempurl}


\bibitem[Vaswani et~al\mbox{.}(2017)]%
        {Vaswani2017Transformers}
\bibfield{author}{\bibinfo{person}{Ashish Vaswani}, \bibinfo{person}{Noam
  Shazeer}, \bibinfo{person}{Niki Parmar}, \bibinfo{person}{Jakob Uszkoreit},
  \bibinfo{person}{Llion Jones}, \bibinfo{person}{Aidan~N. Gomez},
  \bibinfo{person}{Lukasz Kaiser}, {and} \bibinfo{person}{Illia Polosukhin}.}
  \bibinfo{year}{2017}\natexlab{}.
\newblock \showarticletitle{Attention is All you Need}. In
  \bibinfo{booktitle}{\emph{Advances in Neural Information Processing Systems
  30: Annual Conference on Neural Information Processing Systems 2017, December
  4-9, 2017, Long Beach, CA, {USA}}},
  \bibfield{editor}{\bibinfo{person}{Isabelle Guyon}, \bibinfo{person}{Ulrike
  von Luxburg}, \bibinfo{person}{Samy Bengio}, \bibinfo{person}{Hanna~M.
  Wallach}, \bibinfo{person}{Rob Fergus}, \bibinfo{person}{S.~V.~N.
  Vishwanathan}, {and} \bibinfo{person}{Roman Garnett}} (Eds.).
  \bibinfo{pages}{5998--6008}.
\newblock
\urldef\tempurl%
\url{https://proceedings.neurips.cc/paper/2017/hash/3f5ee243547dee91fbd053c1c4a845aa-Abstract.html}
\showURL{%
\tempurl}


\bibitem[Wu et~al\mbox{.}(2022)]%
        {Wu2022InstanceDepPrompt}
\bibfield{author}{\bibinfo{person}{Zhuofeng Wu}, \bibinfo{person}{Sinong Wang},
  \bibinfo{person}{Jiatao Gu}, \bibinfo{person}{Rui Hou},
  \bibinfo{person}{Yuxiao Dong}, \bibinfo{person}{V.~G.~Vinod Vydiswaran},
  {and} \bibinfo{person}{Hao Ma}.} \bibinfo{year}{2022}\natexlab{}.
\newblock \showarticletitle{{IDPG:} An Instance-Dependent Prompt Generation
  Method}.
\newblock \bibinfo{journal}{\emph{CoRR}}  \bibinfo{volume}{abs/2204.04497}
  (\bibinfo{year}{2022}).
\newblock
\urldef\tempurl%
\url{https://doi.org/10.48550/arXiv.2204.04497}
\showDOI{\tempurl}
\showeprint[arXiv]{2204.04497}


\bibitem[Xiong et~al\mbox{.}(2021)]%
        {Xiong2021ANCE}
\bibfield{author}{\bibinfo{person}{Lee Xiong}, \bibinfo{person}{Chenyan Xiong},
  \bibinfo{person}{Ye Li}, \bibinfo{person}{Kwok{-}Fung Tang},
  \bibinfo{person}{Jialin Liu}, \bibinfo{person}{Paul~N. Bennett},
  \bibinfo{person}{Junaid Ahmed}, {and} \bibinfo{person}{Arnold Overwijk}.}
  \bibinfo{year}{2021}\natexlab{}.
\newblock \showarticletitle{Approximate Nearest Neighbor Negative Contrastive
  Learning for Dense Text Retrieval}. In \bibinfo{booktitle}{\emph{9th
  International Conference on Learning Representations, {ICLR} 2021, Virtual
  Event, Austria, May 3-7, 2021}}. \bibinfo{publisher}{OpenReview.net}.
\newblock
\urldef\tempurl%
\url{https://openreview.net/forum?id=zeFrfgyZln}
\showURL{%
\tempurl}


\bibitem[Yang et~al\mbox{.}(2021)]%
        {Yang2021topk}
\bibfield{author}{\bibinfo{person}{Jheng{-}Hong Yang},
  \bibinfo{person}{Xueguang Ma}, {and} \bibinfo{person}{Jimmy Lin}.}
  \bibinfo{year}{2021}\natexlab{}.
\newblock \showarticletitle{Sparsifying Sparse Representations for Passage
  Retrieval by Top-k Masking}.
\newblock \bibinfo{journal}{\emph{CoRR}}  \bibinfo{volume}{abs/2112.09628}
  (\bibinfo{year}{2021}).
\newblock
\showeprint[arXiv]{2112.09628}
\urldef\tempurl%
\url{https://arxiv.org/abs/2112.09628}
\showURL{%
\tempurl}


\bibitem[Yang et~al\mbox{.}(2017)]%
        {Yang2017Anserini}
\bibfield{author}{\bibinfo{person}{Peilin Yang}, \bibinfo{person}{Hui Fang},
  {and} \bibinfo{person}{Jimmy Lin}.} \bibinfo{year}{2017}\natexlab{}.
\newblock \showarticletitle{Anserini: Enabling the Use of Lucene for
  Information Retrieval Research}. In \bibinfo{booktitle}{\emph{Proceedings of
  the 40th International {ACM} {SIGIR} Conference on Research and Development
  in Information Retrieval, Shinjuku, Tokyo, Japan, August 7-11, 2017}},
  \bibfield{editor}{\bibinfo{person}{Noriko Kando}, \bibinfo{person}{Tetsuya
  Sakai}, \bibinfo{person}{Hideo Joho}, \bibinfo{person}{Hang Li},
  \bibinfo{person}{Arjen~P. de~Vries}, {and} \bibinfo{person}{Ryen~W. White}}
  (Eds.). \bibinfo{publisher}{{ACM}}, \bibinfo{pages}{1253--1256}.
\newblock
\urldef\tempurl%
\url{https://doi.org/10.1145/3077136.3080721}
\showDOI{\tempurl}


\bibitem[Zhan et~al\mbox{.}(2021)]%
        {Zhan2021STAR-ADORE}
\bibfield{author}{\bibinfo{person}{Jingtao Zhan}, \bibinfo{person}{Jiaxin Mao},
  \bibinfo{person}{Yiqun Liu}, \bibinfo{person}{Jiafeng Guo},
  \bibinfo{person}{Min Zhang}, {and} \bibinfo{person}{Shaoping Ma}.}
  \bibinfo{year}{2021}\natexlab{}.
\newblock \showarticletitle{Optimizing Dense Retrieval Model Training with Hard
  Negatives}. In \bibinfo{booktitle}{\emph{{SIGIR} '21: The 44th International
  {ACM} {SIGIR} Conference on Research and Development in Information
  Retrieval, Virtual Event, Canada, July 11-15, 2021}},
  \bibfield{editor}{\bibinfo{person}{Fernando Diaz}, \bibinfo{person}{Chirag
  Shah}, \bibinfo{person}{Torsten Suel}, \bibinfo{person}{Pablo Castells},
  \bibinfo{person}{Rosie Jones}, {and} \bibinfo{person}{Tetsuya Sakai}} (Eds.).
  \bibinfo{publisher}{{ACM}}, \bibinfo{pages}{1503--1512}.
\newblock
\urldef\tempurl%
\url{https://doi.org/10.1145/3404835.3462880}
\showDOI{\tempurl}


\bibitem[Zhan et~al\mbox{.}(2022)]%
        {Zhan2022RepCONC}
\bibfield{author}{\bibinfo{person}{Jingtao Zhan}, \bibinfo{person}{Jiaxin Mao},
  \bibinfo{person}{Yiqun Liu}, \bibinfo{person}{Jiafeng Guo},
  \bibinfo{person}{Min Zhang}, {and} \bibinfo{person}{Shaoping Ma}.}
  \bibinfo{year}{2022}\natexlab{}.
\newblock \showarticletitle{Learning Discrete Representations via Constrained
  Clustering for Effective and Efficient Dense Retrieval}. In
  \bibinfo{booktitle}{\emph{{WSDM} '22: The Fifteenth {ACM} International
  Conference on Web Search and Data Mining, Virtual Event / Tempe, AZ, USA,
  February 21 - 25, 2022}}, \bibfield{editor}{\bibinfo{person}{K.~Selcuk
  Candan}, \bibinfo{person}{Huan Liu}, \bibinfo{person}{Leman Akoglu},
  \bibinfo{person}{Xin~Luna Dong}, {and} \bibinfo{person}{Jiliang Tang}}
  (Eds.). \bibinfo{publisher}{{ACM}}, \bibinfo{pages}{1328--1336}.
\newblock
\urldef\tempurl%
\url{https://doi.org/10.1145/3488560.3498443}
\showDOI{\tempurl}


\bibitem[Zhang et~al\mbox{.}(2022)]%
        {Zhang2021AR2}
\bibfield{author}{\bibinfo{person}{Hang Zhang}, \bibinfo{person}{Yeyun Gong},
  \bibinfo{person}{Yelong Shen}, \bibinfo{person}{Jiancheng Lv},
  \bibinfo{person}{Nan Duan}, {and} \bibinfo{person}{Weizhu Chen}.}
  \bibinfo{year}{2022}\natexlab{}.
\newblock \showarticletitle{Adversarial Retriever-Ranker for Dense Text
  Retrieval}. In \bibinfo{booktitle}{\emph{International Conference on Learning
  Representations}}.
\newblock
\urldef\tempurl%
\url{https://openreview.net/forum?id=MR7XubKUFB}
\showURL{%
\tempurl}


\bibitem[Zhou et~al\mbox{.}(2022)]%
        {Zhou2022Hyperlink}
\bibfield{author}{\bibinfo{person}{Jiawei Zhou}, \bibinfo{person}{Xiaoguang
  Li}, \bibinfo{person}{Lifeng Shang}, \bibinfo{person}{Lan Luo},
  \bibinfo{person}{Ke Zhan}, \bibinfo{person}{Enrui Hu}, \bibinfo{person}{Xinyu
  Zhang}, \bibinfo{person}{Hao Jiang}, \bibinfo{person}{Zhao Cao},
  \bibinfo{person}{Fan Yu}, \bibinfo{person}{Xin Jiang}, \bibinfo{person}{Qun
  Liu}, {and} \bibinfo{person}{Lei Chen}.} \bibinfo{year}{2022}\natexlab{}.
\newblock \showarticletitle{Hyperlink-induced Pre-training for Passage
  Retrieval in Open-domain Question Answering}. In
  \bibinfo{booktitle}{\emph{Proceedings of the 60th Annual Meeting of the
  Association for Computational Linguistics (Volume 1: Long Papers), {ACL}
  2022, Dublin, Ireland, May 22-27, 2022}},
  \bibfield{editor}{\bibinfo{person}{Smaranda Muresan},
  \bibinfo{person}{Preslav Nakov}, {and} \bibinfo{person}{Aline Villavicencio}}
  (Eds.). \bibinfo{publisher}{Association for Computational Linguistics},
  \bibinfo{pages}{7135--7146}.
\newblock
\urldef\tempurl%
\url{https://doi.org/10.18653/v1/2022.acl-long.493}
\showDOI{\tempurl}


\end{thebibliography}
